\documentclass[a4paper]{article}

\usepackage{amsmath}
\usepackage{amssymb} 
\usepackage{latexsym} 
\usepackage{picinpar}
\usepackage{curves}
 
\newcommand{\twocond}[2]{\genfrac{}{}{0pt}{}{#1}{#2}}

\def\onetom{\{1,\dots,m\}} 
\def\oneton{\{1,\dots,n\}} 
\def\onetol{\{1,\dots,l\}} 
\newcommand{\vars}{{ vars}} 
\newcommand{\pars}{{ pars}} 
 
\newcommand{\vect}[1]{{\bar{#1}}}  
\newcommand{\dom}{{ dom}} 
 
\def\ftau{f_{\tau_1 \dots \tau_n\rightarrow \tau}} 
\def\ptau{p_{\tau_1 \dots \tau_n}} 
\def\fdep{f_{\mathrm{rel}}}
\def\lar{\leftarrow}
\def\rar{\rightarrow}
\newcommand{\nt}[1]{\mbox{`}#1\mbox{'}}
\newcommand{\vt}{\Gamma}
\newcommand{\set}[1]{\{#1\}}
\newcommand{\bigset}[1]{\left\{#1\right\}}
\def\appl#1#2{#2 #1}
\def\subone{\triangleleft}

\def\sub{\triangleleft^*} 
\def\rec{\bowtie}
\def\Rec{\mathop{\Join}}

\def\nrs{\mathrel{\mathord{\triangleleft\hspace{-0.25em}\triangleleft}}}
\def\Nrs{\mathop{\mathord{\triangleleft\hspace{-0.25em}\triangleleft}}}
\def\notsub{\mathbin{\mathord{\not\!\sub}}}

\newcommand{\describes}{\propto}
\newcommand{\sem}[1]{\mathord{[\hspace{-0.15em}[}#1\mathord{]\hspace{-0.15em}]}}
\newcommand{\renamed}[1]{\ll_{#1}}
\newcommand{\theory}{AC+}
\newcommand{\absleq}{\mathrel{\leq_\mathrm{\theory}}}

\def\newplaintheorem#1#2#3{%
\newtheorem{#1plain}[#3]{#2}%[section]
\newenvironment{#1}{\begin{#1plain}\rm}{\end{#1plain}}}

\newplaintheorem{ex}{Example}{theo}
\newplaintheorem{defi}{Definition}{theo}
\newplaintheorem{coro}{Corollary}{theo}
\newplaintheorem{rmtheorem}{Theorem}{theo}
\newplaintheorem{rmlemma}{Lemma}{theo}
\newplaintheorem{propo}{Proposition}{theo}

\newenvironment{pf}
 {{\normalfont \noindent {\scshape Proof.}}% 
   \setlength{\parindent}{0pt}%
   \setlength{\parskip}{1.5ex plus 0.5ex minus 1.0ex}% 
}
 {\hspace*{1em}\hfill$\Box$
\vspace{1ex}}

\begin{document} 
%\mainmatter 
 
\bibliographystyle{alpha} 
 
\title{Analysis of Polymorphically Typed 
Logic Programs Using ACI-Unification}

\author{Jan--Georg Smaus\thanks{
CWI, Amsterdam, The Netherlands,
\texttt{jan.smaus@cwi.nl}}}

\maketitle 

\begin{abstract} 
Analysis of (partial) groundness is an important application of
abstract interpretation. There are several proposals for improving the
precision of such an analysis by exploiting type information, icluding
our own work with Hill and King~\cite{SHK00}, where we had shown how
the information present in the type declarations of a program can be
used to characterise the degree of instantiation of a term in a
precise and yet inherently finite way. This approach worked for {\em
polymorphically} typed programs as in G\"odel or HAL. Here, we recast
this approach following Codish, Lagoon and
Stuckey~\cite{CL00,LS01}. To formalise which properties of
terms we want to characterise, we use {\em labelling functions}, which
are functions that extract subterms from a term along certain
paths. An {\em abstract term} collects the results of all labelling
functions of a term. For the analysis, programs are executed on 
abstract terms instead of the concrete ones, and usual 
unification is replaced by unification modulo an equality theory which
includes the well-known ACI-theory. Thus we
generalise~\cite{CL00,LS01} w.r.t.~the type systems considered and
relate those two works.
\end{abstract} 

\section{Introduction}\label{intro-sec}

Analysing logic programs for properties such as sharing and (partial)
groundness is important in compiler optimisations and
program development tools. Programs are usually analysed using
abstract interpretation~\cite{CC77}. In this paper, we consider in
particular the framework of {\em abstract compilation}~\cite{CD95}: 
a program is abstracted by replacing each unification with
an abstract counterpart, and then the abstract program is evaluated
just like a concrete program.

It has been recognised for some time that abstract interpretation can
be used for type analysis, and conversely, that type
information available a priori can improve the precision of other
analyses \cite{BM95,CD94,CL00,GdW94,JB92,HCC93}. For example, being
able to say that $\tt [1,X]$ is a list skeleton with possibly 
uninstantiated elements is more precise than only being able to
distinguish a ground from a possibly non-ground term. Underlying all
those works is a {\em descriptive} view of types: types are not part
of the programming language (in particular, no program is rejected
for not being ``well-typed''), but rather introduced to analyse an
arbitrary, say Prolog, program. In such approaches, it is natural that 
there is no sharp line between {\em type} analysis and {\em mode} 
(groundness, instantiation) analysis. For example, saying that a term 
is a list has two aspects: it is a list as opposed to, say, an
integer; it is a list, as opposed to an uninstantiated variable. 

Underlying this paper is a {\em prescriptive} view of types, i.e.,
types are a part of the programming language. We
analyse programs written in typed logic programming languages such as
G\"odel~\cite{goedel}, HAL~\cite{DGHMS99}, or
Mercury~\cite{mercury}. This implies that the types do not have to be
analysed since they are given beforehand by declarations or
inference. In particular, unlike e.g.~\cite{CL00}, we do not have to
deal with ``ill-typed'' terms such as $[1|2]$, since these can never
occur. 

We are aware of only two other works following
this view: our own~\cite{SHK00} and the recent work by Lagoon and
Stuckey~\cite{LS01}. This paper is a synthesis of those two works
and~\cite{CL00}, which, although designed for a
descriptive view of typing, can be adapted to prescriptive
typing.\footnote{The journal article~\cite{CL00} is based on an
earlier article~\cite{CL96}. However there are some interesting
differences, and therefore we will also sometimes refer to the earlier 
article.}
The generalisation w.r.t.~\cite{CL00,LS01} concerns {\em polymorphism},
which is disregarded in~\cite{LS01} and considered in~\cite{CL00} only
in a restricted form. We recast our own previous work using some 
aspects of their formalisms. In particular, the use
of the notions of {\em grammar} and {\em variables labelling
non-terminals}~\cite{LS01} should improve the understanding of what 
properties of terms our analysis captures, whereas the use of
ACI-unification~\cite{CL00} may provide the basis for an
implementation using well-studied algorithms. Also, we hope that our
work will prove to be applicable to analyses previously not envisaged by us, 
such as {\em sharing} analysis~\cite{LS01}.

In the intuitive explanations that follow, we refer to a set of
possible characterisations of the instantiation of a term as 
{\em abstract domain}.

The standard example to illustrate the benefits of an instantiation
analysis using types is the ubiquitous \texttt{APPEND} program. For
example, for the query 
$\tt append([A],[B],C)$, a typed analysis is able to infer that
any answer substitution will bind $\tt C$ to a list
skeleton. However, this example is unfit to explain the advance of
this paper over previous works. 

We therefore give another example.  A {\em table} is a data structure
containing an ordered collection of nodes, each of which has two
components, a key of type $\mathtt{string}$\footnote{
We abbreviate $\mathtt{string}$ by $\mathtt{str}$ and 
$\mathtt{integer}$ by $\mathtt{int}$.}, 
and a value,  of arbitrary type.
That is to say, the type constructor $\mathtt{table}$ is pa\-ra\-met\-rised by
the type of the values. For any type $\tau$, 
$\mathtt{table}(\tau)$ is the type of tables whose values have type 
$\tau$. Tables are implemented in G\"odel
  as an AVL-tree~\cite{vE81}:  a non-leaf node has a 
{\em key} argument, a {\em value} argument, arguments for the left and
  right subtrees, and an argument which represents balancing
  information. For a term of type $\mathtt{table}(\tau)$, our abstract 
domain characterises the instantiation of all key arguments, all value
arguments, and all the arguments representing the balancing
information. 

The characterisation of the instantiation of the value arguments
depends on $\tau$. Hence, our analysis supports parametric
polymorphism. In devising an analysis for polymorphically typed
programs, there are two main problems: the construction of an abstract 
domain for $\mathtt{table}(\tau)$ should be truly parametric in
$\tau$, and the abstract domains should be finite for a given program
and query. We only briefly illustrate what these points mean
here. Explaining why these requirements are non-trivial is
technically too involved for this introduction. 

The statement that the construction of an abstract domain for
$\mathtt{table}(\tau)$ is truly parametric in $\tau$, means,
for example, that the abstract domain for $\mathtt{table}(\mathtt{str})$
relates to $\mathtt{str}$ in exactly the same way as the abstract
domain for $\mathtt{table}(\mathtt{int})$ relates to 
\addtocounter{footnote}{-1}
$\mathtt{int}$\footnotemark. This implies that the abstraction of a
table can be defined in a generic way.

Lagoon and Stuckey formalise types as {\em regular tree
grammars}. Each type is identified with a non-terminal in
the grammar, and it is assumed that there are only finitely many
types. Finiteness is crucial for the termination of an analysis. When
we extend this approach to polymorphism, finiteness becomes a
problem, since there are infinitely many types, 
e.g.~$\tt list(int),\ list(list(int(int))), \dots$. Nevertheless,
under certain conditions, it can be ensured that for a given query and
program, there are only finitely many types. Note that this is in
contrast to~\cite{CL00} where it is proposed that termination of
analyses of polymorphic programs should be enforced by imposing an
ad-hoc bound on the depth of types.

%Although we follow the formalism of Lagoon and Stuckey in many ways,
%there are some differences that we want to note beforehand. They claim
%that their approach works for prescriptively typed programs, say
%G\"odel programs. But then in the formal part, the ``typedness'' of a
%program only means that each program variable is annotated with a
%grammar defining an expected type for this variable. We believe that
%prescriptive typing goes further than that. It must become clear that
%some annotations are legal and some are not, and for some programs no
%legal annotation exists. 

%This has a number of consequences. In particular, in the formalism
%of~\cite{LS01}, a unification constraint $\tt X = Y$ in a program
%means that the intersection of the two types corresponding to $\tt X$
%and $\tt Y$ must be computed. We believe that for prescriptively typed 
%programs, this should not be necessary, as the types of $\tt X$
%and $\tt Y$ must be identical anyway for the program to be
%well-typed. 

The rest of this paper is organised as follows.
The next section provides some preliminaries. 
In Sec.~\ref{structure-sec}, following~\cite{LS01}, 
we show how the type of a term gives rise to characterising 
its degree of instantiation in a structured way.
In Sec.~\ref{abstract-terms-sec}, following~\cite{CL00}, 
we define abstract terms based on the ACI1 equality theory.
In Sec.~\ref{alpha-label-relation-sec}, we formalise how abstract
terms capture the degree of instantiation of concrete terms, thereby
linking the two preceding sections, and also linking~\cite{LS01}
with~\cite{CL00}. 
Section~\ref{analysis-sec} lifts the abstraction of terms to an
abstraction of programs, and relates the semantics of a concrete
program and its abstraction in the sense of abstract interpretation. 
Section~\ref{implementation-sec} makes some comments on a possible
future implementation, and Sec.~\ref{discussion-sec} discusses our
results.

\section{Preliminaries}\label{prelim-sec}
The reader is assumed to be familiar with the basics of logic
programming~\cite{L87}. We use 
a type system for logic programs with parametric
polymorphism~\cite{DGHMS99,goedel,mercury}. 

Let $\mathcal{K}$ be a finite set of (type) {\bf constructors},
each
$c\in\mathcal{K}$ with an arity $n\geq 0$ associated (by writing $c/n$),
and $\mathcal U$ be a denumerable set of {\bf parameters}.
The set of types is the term structure $\mathcal{T(K,U)}$. 
A {\bf type substitution} is an idempotent mapping
from parameters to types which is the identity almost everywhere.
We define the order $\prec$ on
types as the order induced by some (for example lexicographical) order on
constructor and parameter symbols, where parameter symbols come before
constructor symbols. 

The set of parameters in a syntactic object $o$ is denoted by $\pars(o)$.
Parameters are denoted by $u,v$, in concrete
examples by $\mathtt{U, V}$.
A tuple of {\em distinct} parameters ordered
with respect to~$\prec$ is denoted by $\bar{u}, \bar{v}$. 

Let $\mathcal V$ be a denumerable set of {\bf variables}.
The set of variables in a syntactic object $o$ is
denoted by $\vars(o)$.
Variables are denoted by $x, y$, in concrete
examples by $\mathtt{X, Y}$.
A tuple of {\em distinct} variables is denoted by 
$\vect{x}, \vect{y}$.  

A {\bf variable typing} is a mapping from a finite subset of 
$\mathcal V$ to $\mathcal{T(K,U)}$, written as
$\{x_1:\tau_1,\dots,x_n:\tau_n\}$.

Let $\mathcal F$ (resp.~$\mathcal P$) 
be a finite set of {\bf function} (resp.~{\bf predicate}) symbols, each
with an arity and a {\bf declared type} associated with it, 
such that:
for each $f \in \mathcal F$, the declared type has the form
$(\tau_1,\dots,\tau_n,\tau)$, where $n$ is the arity of $f$,
$(\tau_1,\dots,\tau_n,\tau)\in \mathcal{T(K,U)}^{n+1}$,
and $\tau$ 
satisfies the {\em transparency condition} \cite{HT92}:\label{transparency}
$\pars(\tau_1,\dots,\tau_n) \subseteq \pars(\tau)$;
for each $p \in \mathcal P$, the declared type has the form
$(\tau_1,\dots,\tau_n)$, 
where $n$ is the arity of $p$ and
$(\tau_1,\dots,\tau_n)\in \mathcal{T(K,U)}^n$. 
We often indicate the declared types by writing 
$\ftau$ and $\ptau$.
%We assume that there is a special predicate symbol 
%$=_{u,u}$
%where $u\in \mathcal U$.

Throughout this paper, we assume $\mathcal K$, $\mathcal F$, and 
$\mathcal P$ arbitrary but fixed.
The {\bf typed language}, i.e.\ a language of terms, atoms etc.\ based
on $\mathcal K$, $\mathcal F$, and $\mathcal P$, is defined by the
rules in Table \ref{rules-tab}. All objects are defined relative to 
a variable typing $\vt$, and $\_\vdash\dots$ stands for ``there exists
$\vt$ such that $\vt\vdash\dots$''. 
Actually, we will rarely refer to the type system explicitly, but it
should be noted that any objects we will come across in the context of 
analysing a typed program will be correctly typed according to those
rules. This is guaranteed because typed programs have the 
{\em subject reduction} property~\cite{HT92}.

\begin{table}[t]
\caption{Rules defining a typed language ($\Theta$ is a type
substitution) \label{rules-tab}}
\begin{center}
%\small
\begin{tabular}{rl}
{\em (Var)} &
$\{x:\tau,\dots\}\vdash x:\tau$\\[2ex]
{\em (Func)} &
\Large
$\frac%
  {\vt\vdash t_1:\tau_1\Theta\ \cdots \ \vt\vdash t_n:\tau_n\Theta}%
  {\vt\vdash\ftau(t_1,\dots,t_n):\tau\Theta}$ 
%& $\Theta$ is a type substitution
\\[2ex]
{\em (Atom)} &
\Large
$\frac%
  {\vt\vdash t_1:\tau_1\Theta\ \cdots \ \vt\vdash t_n:\tau_n\Theta}%
  {\vt\vdash\ptau(t_1,\dots,t_n)\; \mathit{Atom}}$ 
%& $\Theta$ is a type substitution
\\[2ex]
{\em (Head)} &
\Large
$\frac%
  {\vt\vdash t_1:\tau_1\ \cdots \ \vt\vdash t_n:\tau_n}%
  {\vt\vdash\ptau(t_1,\dots,t_n)\; \mathit{Head}}$ 
%\\[2ex]
\end{tabular}
\begin{tabular}{rl}
{\em (Query)} &
\Large
$\frac%
  {\vt\vdash A_1\; \mathit{Atom}\ \cdots \  \vt\vdash A_n\; \mathit{Atom}}%
  {\vt\vdash A_1,\dots,A_n\; \mathit{Query}}$ \\[2ex]
{\em (Clause)} & 
\Large
$\frac%
  {\vt\vdash A\; \mathit{Head} \quad \vt\vdash Q\; Query}%
  {\vt\vdash A \leftarrow Q\; \mathit{Clause}}$ \\[2ex]
{\em (Program)} &
\Large
$\frac%
  {\_\vdash C_1\; \mathit{Clause} \ \cdots \ \_\vdash C_n\; \mathit{Clause}}%
  {\_\vdash \{C_1,\dots,C_n\}\; \mathit{Program}}$ 
%\\[2ex]
%{\em (Queryset)} &
%\Large
%$\frac%
%  {\_\vdash Q_1\; \mathit{Query}\ \cdots \  \_\vdash Q_n\; \mathit{Query}}%
%  {\_\vdash \{Q_1,\dots, Q_n\} \; \mathit{Queryset}}$ 
\end{tabular}
\end{center}
\end{table}

%\begin{defi}\label{substitution-def}
%If 
%$\vt\vdash x_1\!=\!t_1, \dots, x_n\! =\! t_n\ Query$
%where $x_1,\dots,x_n$ are distinct variables and for
%each $i\in \onetom$, $t_i$ is a term distinct from $x_i$, then 
%$(\{ x_1/t_1,\dots,x_n/t_n\}, \vt)$ is a 
%{\bf typed (term) substitution}. 
%\end{defi}

%The restriction of a 
%substitution $\theta$ to the variables in a syntactic object $o$ is
%denoted as $\theta \restr{o}$, and analogously for type 
%substitutions.
%The relation symbol of an atom $a$ is denoted by $Rel(a)$.

%Programs are assumed to be in {\em canonical form}: 
%a {\bf canonical literal}
%is a unification constraint (equation) of the
%    form $x = y$ or 
%$x = f(\bar{y})$, or an
%    atom $p(\bar{y})$.
%A {\bf canonical query} is a sequence of canonical literals. 
%A {\bf canonical clause} has the form 
%$p(\bar{y}) \lar Q$ where $Q$ is a canonical query.
%A program is {\bf canonical} if all of its clauses are. 

Concerning semantics, we entirely follow~\cite{CL00}.
The set of atoms is denoted by $\mathcal{B}$, and 
elements of $2^\mathcal{B}$ are called 
{\bf interpretations}. 
For a syntactic object $o$ and a set of objects $I$, we denote by
$\langle C_1,\dots,C_n\rangle\renamed{o} I$ that 
$C_1,\dots,C_n$ are elements of $I$ renamed apart from $o$ and from
each other.
So the analysis we shall propose is generic and independent of any particular
(say top-down or bottom-up) concrete semantics, but examples will be
given using the $s$-semantics, i.e.~the semantics based on the
non-ground $T_P$-operator, defined as follows:
\begin{align*}
T_P(I) = 
\set{H\theta \mid & 
C = H \lar B_1,\dots,B_n \in P,
\langle A_1,\dots,A_n \rangle \renamed{C} I,\\[-1ex]
&
\theta = 
MGU(\langle B_1,\dots,B_n\rangle,\langle A_1,\dots,A_n\rangle)}.
\end{align*}
We denote by $\sem{P}_s$ the least fixpoint of $T_P$.

We denote by $t_1\leq t_2$ that $t_1$ is an instance if $t_2$.
The domain of a substitution $\theta$ is denoted as $\dom(\theta)$.

\section{The Structure of Terms and Types}\label{structure-sec}
In this section, we show how the type of a term gives rise to a
certain way of characterising its structure, and in particular, how
its degree of instantiation can be characterised in a structured way. 
We alternate between recalling the
formalism of~\cite{LS01}, and adapting it to polymorphic types,
thereby linking to~\cite{SHK00}. 

\subsection{Regular Types~\cite{LS01}}\label{regular-subsec}

\begin{defi}\label{automaton-def}
A {\bf top-down deterministic finite tree automaton} (top-down DFTA)
is a tuple 
$\mathcal{A} = \langle q_0, Q, \Sigma, \Delta\rangle$, where $Q$ is a
set of states, $q_0 \in Q$ is an initial state and $\Delta$ is a set
of transition rules of the form
$q(f(x_1,\dots,x_n)) \rightarrow f(q_1(x_1),\dots,q_n(x_n))$, such
that no two rules have the same left-hand side.
\end{defi}

Top-down DFTA's accept the class of languages called {\em regular
types}. 

\begin{defi}\label{grammar-def}
A {\bf regular tree grammar} is a tuple 
$\mathcal{G} = \langle S, W, \Sigma, \Delta \rangle$, where
$W$ is a finite set of non-terminal symbols, 
$S\in W$ is a starting non-terminal, 
$\Delta$ is a set of productions in the form
$X \rightarrow f(Y_1,\dots,Y_n)$ s.t.~$X,Y_1,\dots,Y_n\in W$ and 
$f/n \in \Sigma$. 
A regular tree grammar is {\bf deterministic} if for any 
non-terminal $X$ and any two productions
$X \rightarrow f(Y_1,\dots,Y_n)$ and 
$X \rightarrow g(Z_1,\dots,Z_m)$, we have 
$f/n \neq g/m$.
\end{defi}

It has been pointed out that the two formalisms above define the same
class of languages. Transitions of the automaton can be converted to
grammar productions and vice versa by identifying each non-terminal
symbol with a corresponding state of the automaton. 

\begin{ex}\label{same-class-ex}
The DFTA 
$
\langle q_L, \set{q_L, q_E}, 
\set{\mathtt{nil}/0, \mathtt{cons}/2, \mathtt{a}/0, \mathtt{b}/0},
\Delta \rangle
$,
where 
\[
\Delta = \set{%
q_L(\mathtt{nil}) \rar\mathtt{nil},\;
q_L(\mathtt{cons}(x,y)) \rar\mathtt{cons}(q_E(x),q_L(y)),\;
q_E(\mathtt{a}) \rar\mathtt{a},\;\allowbreak
q_E(\mathtt{b}) \rar\mathtt{b}
},
\] 
accepts ground lists of $\mathtt{a}$'s and $\mathtt{b}$'s.

The grammar 
$L \rar\mathtt{nil} | \mathtt{cons}(E,L),\;
E \rar\mathtt{a} | \mathtt{b}$ 
defines the same language.
\end{ex}

The equivalence of the two formalisms allows us to represent derivations of
the grammar as transitions of a {\em deterministic} rewriting
system. These transitions have the form 
$N(f(t_1,\dots,t_n)) \rar f(N_1(t_1),\dots,N_n(t_n))$, where 
$N \rightarrow f(N_1,\dots,N_n)$ is a production of the grammar 
($n \geq 0$). Using this notation, we say that the grammar
$\mathcal{G}= \langle S, W, \Sigma, \Delta \rangle$ 
{\bf accepts} a term $t$ if $S(t) \rar^* t$. Sometimes we are
interested in a particular segment of a single path in a derivation 
tree starting from root $S$ and reaching a non-terminal $N$ with a
subterm $t'$ of $t$, i.e., in derivations
$S(t) \rar^* s[N(t')]$, where the notation $s[N(t')]$ means that $s$
has $N(t')$ as a subterm. Abusing notation, we write
$S(t) \rar^* N(t')$ in this case. 

\begin{ex}\label{path-ex}
Given the grammar in Ex.~\ref{same-class-ex},
we have 
\[
L(\mathtt{cons}(\mathtt{a},\mathtt{nil})) \rar
\mathtt{cons}(E(\mathtt{a}), L(\mathtt{nil})) \rar
\mathtt{cons}(\mathtt{a}, L(\mathtt{nil})) \rar
\mathtt{cons}(\mathtt{a}, \mathtt{nil}).
\] 
We also write 
$L(\mathtt{cons}(\mathtt{a},\mathtt{nil})) \rar^* E(\mathtt{a})$ and
$L(\mathtt{cons}(\mathtt{a},\mathtt{nil})) \rar^* L(\mathtt{nil})$,
using the above notation.

This notation can also be applied to non-ground terms. For example, we 
have
$L(\mathtt{cons}(\mathtt{a},\mathtt{Y})) \rar^* E(\mathtt{a})$ and
$L(\mathtt{cons}(\mathtt{X},\mathtt{Y})) \rar^* L(\mathtt{Y})$. 
\end{ex}

\setlength{\unitlength}{3mm}

\begin{figwindow}[0,r,%
{\mbox{\hspace{1em}\begin{picture}(10,4)%
{\linethickness{0.2\unitlength}
\put(1,0){\framebox(2,3){$L$}}}%
\curve(3,2, 4,2.5, 4,4, 2.5,4, 2,3)%
\put(7,0){\framebox(2,3){$E$}}%
\put(3,1){\vector(1,0){4}}%
\put(2,3){\vector(0,-1){0}}%
\end{picture}}},%
{Type graph \label{one-type-graph-fig}}]
It is also convenient to depict a grammar graphically as a 
{\em type graph}, defined previously as a graph whose nodes are
labelled with types or functions~\cite{HCC93}. We simplify that
definition by leaving out the function nodes. Thus a type graph for 
$\mathcal{G} = \langle S, W, \Sigma, \Delta \rangle$ is
a directed graph whose nodes are labelled by non-terminals, and there
is an edge from $N$ to $N'$ if and only if there is a production 
$N \rar f(\dots,N',\dots)$ is $\mathcal{G}$. We call the node labelled
$S$ the {\bf starting node}. Figure~\ref{one-type-graph-fig} shows the type
graph for Ex.~\ref{same-class-ex}. 
\end{figwindow}

\subsection{Regular Types and Polymorphism}
\label{regular-polymorphic-subsec}
Converting the type declarations of a typed language such as Mercury
into grammar rules has been considered 
straightforward~\cite[footnote 1]{LS01}. This
seems justified, albeit only in the absence of polymorphism. Since
$\mathcal K$ is a {\em finite} set of type constants, we can identify
each type constant with a non-terminal, and each function 
$\ftau \in \mathcal{F}$ is translated into a production
$\tau \rar f(\tau_1,\dots,\tau_n)$.
 In that way, each type (constant)
corresponds to a grammar with that type as starting
non-terminal.\footnote{
Although one might be
confused by the fact that~\cite{LS01} also says that there is a
grammar for each {\em program variable}, but this is simply a matter
of renaming.}  
Table~\ref{formalisms-tab} summarises the correspondences between
the four formalisms we effectively identify in this paper.

\begin{table}[t]
\begin{center}
\begin{tabular}{llll}
Automata                                & Grammars     & 
Types                                   & Graphs \\\hline
state                                   & non-terminal\hspace{1em} & 
type                                    & node \\
transition rule\hspace{1em}             & production   & 
$\approx$ type declarations\hspace{1em} & edge
\end{tabular}
\end{center}
\caption{Correspondences between formalisms\label{formalisms-tab}}
\end{table}

Note that in Sec.~\ref{prelim-sec}, we have specified that each $f$
has {\em exactly} one declaration; in other words, there is no
overloading. This is a sufficient condition for the grammar to be
deterministic. One could allow some overloading, by specifying: if 
$\ftau\in \mathcal{F}$ and  
$f_{\sigma_1,\dots,\sigma_m\rar\sigma}\in \mathcal{F}$ and
$\ftau \neq f_{\sigma_1,\dots,\sigma_m\rar\sigma}$, then either 
$\tau\neq\sigma$, or $n\neq m$. This would strictly include the
overloading allowed in G\"odel. We prefer however to disallow
overloading to avoid unnecessary complication.  

We now give a pseudo-definition of a grammar corresponding to a
polymorphic type --- ``pseudo'' because the set of non-terminals may
be infinite. It is trivial to generalise the definition of a grammar
to infinite sets of non-terminals, but in the end, it is not
desirable, since it would undermine our goal of characterising
(approximating) properties of a term of arbitrary type in a finite
way. We will later impose a condition to ensure finiteness.

\begin{defi}\label{corresponding-grammar-def}
Consider a typed language given by $\mathcal{K, F}$
 and a type $\phi$. The 
{\bf grammar corresponding to $\phi$}, denoted 
$\mathcal{G}(\phi)$, 
is the grammar 
$\langle \nt{\phi}, W, \mathcal{F}, \Delta \rangle$, 
where $W$ is inductively defined as follows
\begin{itemize}
\item
$\nt{\phi} \in W$,
\item
$\ftau\in\mathcal{F}$ and
$\nt{\tau\Theta} \in W$ for some type substitution $\Theta$ implies\\
$\nt{\tau_1\Theta},\dots,\nt{\tau_n\Theta} \in W$,
\end{itemize}
and $\Delta = 
\set{\nt{\tau\Theta} \rar
\ftau(\nt{\tau_1\Theta},\dots,\nt{\tau_n\Theta})
\mid \tau\Theta \in W}$.
\end{defi}

We put types in quotation marks to indicate 
that when looking at the grammar, types are just non-terminal symbols. 
Type graphs are defined as before.
In Fig.~\ref{type-graphs-fig}, we give some type graphs to
which we will refer frequently.

\setlength{\unitlength}{2.3mm}
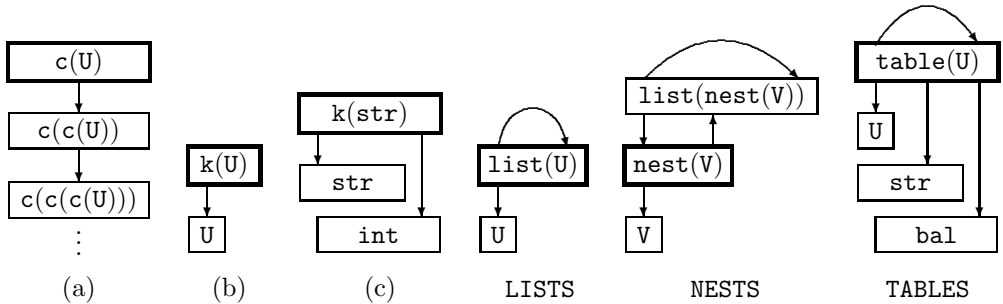
\begin{figure}[t]
\begin{picture}(8,12)
\put(0,0){\makebox(8,2){$\vdots$}}
\put(0,2){\framebox(8,2){$\mathtt{c(c(c(U)))}$}}
\put(4,6){\vector(0,-1){2}}
\put(0,6){\framebox(8,2){$\mathtt{c(c(U))}$}}
\put(4,10){\vector(0,-1){2}}
{\linethickness{0.2\unitlength}
\put(0,10){\framebox(8,2){$\mathtt{c(U)}$}}
}
\end{picture}
\hfill
\begin{picture}(4,6)
\put(0,0){\framebox(2,2){$\tt U$}}
\put(1,4){\vector(0,-1){2}}
{\linethickness{0.2\unitlength}
\put(0,4){\framebox(4,2){$\mathtt{k(U)}$}}
}
\end{picture}
\hfill
\begin{picture}(8,9)
\put(1,0){\framebox(7,2){$\mathtt{int}$}}
\put(0,3){\framebox(6,2){$\mathtt{str}$}}
\put(1,7){\vector(0,-1){2}}
\put(7,7){\vector(0,-1){5}}
{\linethickness{0.2\unitlength}
\put(0,7){\framebox(8,2){$\mathtt{k(str)}$}}
}
\end{picture}
\hfill
\begin{picture}(6,7)
\put(0,0){\framebox(2,2){$\tt U$}}
\put(1,4){\vector(0,-1){2}}
{\linethickness{0.2\unitlength}
\put(0,4){\framebox(6,2){$\mathtt{list(U)}$}}
}
\curve(1,6, 2,8, 4,8, 5,6)
\put(5,6){\vector(1,-4){0}}
\end{picture}
\hfill
\begin{picture}(11,11)
\put(0,0){\framebox(2,2){$\tt V$}}
\put(1,4){\vector(0,-1){2}}
{\linethickness{0.2\unitlength}
\put(0,4){\framebox(6,2){$\mathtt{nest(V)}$}}
}
\put(1,8){\vector(0,-1){2}}
\put(5,6){\vector(0,1){2}}
\put(0,8){\framebox(11,2){$\mathtt{list(nest(V))}$}}
\curve(1,10, 4,12, 7,12, 10,10)
\put(10,10){\vector(1,-1){0}}
\end{picture}
\hfill
\begin{picture}(8,13)
\put(1,0){\framebox(7,2){$\mathtt{bal}$}}
\put(0,3){\framebox(6,2){$\mathtt{str}$}}
\put(0,6){\framebox(2,2){$\tt U$}}
\put(1,10){\vector(0,-1){2}}
\put(4,10){\vector(0,-1){5}}
\put(7,10){\vector(0,-1){8}}
{\linethickness{0.2\unitlength}
\put(0,10){\framebox(8,2){$\mathtt{table(U)}$}}
}
\curve(1,12, 3,14, 5,14, 6.9,12.1)
\put(6.9,12.1){\vector(2,-3){0}}
\end{picture}
%\hfill
%\begin{picture}(6,10)
%{\linethickness{0.2\unitlength}
%\put(0,0){\framebox(6,2){$\mathtt{elist(U)}$}}}
%\put(0,4){\framebox(2,2){$\tt U$}}
%\put(5,2){\vector(0,1){6}}
%\put(3,8){\vector(0,-1){6}}
%\put(1,2){\vector(0,1){2}}
%\put(1,8){\vector(0,-1){2}}
%\put(0,8){\framebox(6,2){$\mathtt{olist(U)}$}}
%\end{picture}

\begin{picture}(8,2)
\put(0,0){\makebox(8,0){(a)}}
\end{picture}
\hfill
\begin{picture}(5,2)
\put(0,0){\makebox(5,0){(b)}}
\end{picture}
\hfill
\begin{picture}(8,2)
\put(0,0){\makebox(8,0){(c)}}
\end{picture}
\hfill
\begin{picture}(6,2)
\put(0,0){\makebox(6,0){\texttt{LISTS}}}
\end{picture}
\hfill
\begin{picture}(11,2)
\put(0,0){\makebox(11,0){\texttt{NESTS}}}
\end{picture}
\hfill
\begin{picture}(8,2)
\put(0,0){\makebox(8,0){\texttt{TABLES}}}
\end{picture}
\caption{Some type graphs, with starting node highlighted\label{type-graphs-fig}}
\end{figure} 

It is also useful to have names and a notation for the relations
holding between the types in a type graph.

\begin{defi} \label{subterm-type-def}
A type $\sigma$ is a 
{\bf direct subterm type of} $\phi$ (denoted as $\sigma \subone \phi$) if
 there is $\ftau \in \mathcal{F}$ and a type substitution $\Theta$ such
 that $\appl{\Theta}{\tau} = \phi$ and $\appl{\Theta}{\tau_i} = \sigma$ for some $i \in
 \oneton$.  The transitive, reflexive closure of $\subone$ is
 denoted as $\sub$. 
If $\sigma \sub  \phi$, then $\sigma$ is a 
{\bf subterm type of} $\phi$. 
\end{defi}

We now discuss two problems related to the generalisation to
polymorphism, including that of finiteness mentioned above. 

\begin{ex}\label{type-graphs-ex}
Whenever we give a particular typed language, $\mathcal K$ is given
implicitly as the set of all type constructors occurring in the type
subscripts in $\mathcal F$.

One would hope that even if a typed language contains an
infinite set of types, the type graph taking a fixed type as starting
node should be finite. However, consider 
$\mathcal{F} =
\set{\mathtt{f_{c(c(U)) \rar c(U)}}}$. The type graph of
$\mathtt{c(U)}$ is shown in
Fig.~\ref{type-graphs-fig} (a). As can be seen, it is infinite.
\end{ex}

We impose the following restriction on any typed language to ensure finiteness.

 \vspace{0.5em}   

\noindent {\bf Reflexive Condition:}  For all $c \in \mathcal{K}$ and 
types $\sigma = c(\bar{\sigma}), \tau = c(\bar{\tau})$, 
if $\sigma \sub\tau$, 
then $\sigma$ is a sub``term'' (in the syntactic sense) of $\tau$.

\vspace{0.5em}   

Clearly, this condition is violated by the example above, where 
 $\mathtt{c(c(U)) \subone c(U)}$. With this condition in place, it 
is easy to see that any type graph for a given starting node is
finite. 

In the introduction we mentioned another problem, namely that the construction
of abstract domains should be ``truly parametric''. 

\begin{ex}\label{truly-parametric-ex}
Consider
$\mathcal{F} = \set{
\mathtt{f_{U\rar k(U)}},
\mathtt{g_{int \rar k(str)}}
}$.
Figure~\ref{type-graphs-fig} (b) shows the type
graph for $\mathtt{k(U)}$. Essentially, the type graph for any
instance $\mathtt{k}(\tau)$ is obtained by replacing the node
$\mathtt{k(U)}$ with $\mathtt{k}(\tau)$ and the node 
$\tt U$ with the type graph for $\tau$. However, there is one
exception to this: if $\tau = \mathtt{str}$, then the type graph is
the one shown in Fig.~\ref{type-graphs-fig} (c). For this
example, it would clearly be wrong to say that 
``$\mathtt{k(U)}$ relates to $\tt U$ in the same way as 
$\mathtt{k(str)}$ relates to $\mathtt{str}$''.
\end{ex}

Again, we rule out this anomaly. First we define:

\begin{defi}\label{flat-def}
A {\bf flat type} is a type of the form $c(\bar{u})$, where 
$c \in \mathcal{K}$.
\end{defi}

We now impose the following condition on any typed language.

\vspace{0.5em}

\noindent {\bf Flat Range Condition:} For all $\ftau \in \mathcal{F}$,
$\tau$ is a flat type.

\vspace{0.5em}   

In Mercury (and also in {\em functional} languages such as ML or
Haskell), this condition is enforced by the
syntax. In G\"odel, it is possible to violate the
condition, but this can be regarded as an artefact of that syntax. 

Thus we assume from now on that any typed language we consider meets
the two conditions above.

\subsection{Labelling~\cite{LS01}}\label{labelling-subsec}
Labellings can be used to characterise the degree of
instantiation of a term taking its type into account, i.e.,
analyse a term on a {\em per-role} basis~\cite{LS01}. 

\begin{defi}\label{labelling-def}
A variable $x$ in a term $t$ labels a non-terminal $N$ of a grammar
$\mathcal{G}$ if $S(t) \rar^* N(x)$, where $S$ is the starting
non-terminal of $\mathcal{G}$. 

We denote by $\zeta(S,N,t)$ the function
which returns the set of
variables $x$ such that $S(t) \rar^* N(x)$ (one could also write
$\zeta(\mathcal{G},N,t)$~\cite{LS01}).
\end{defi}

\begin{ex}\label{labelling-ex}
The grammar
$LL \rar\mathtt{nil} | \mathtt{cons}(L,LL),\;
L \rar\mathtt{nil} |\mathtt{cons}(E,L),\;\allowbreak
E \rar\mathtt{a} | \mathtt{b}$ 
accepts ground lists of lists of 
$\mathtt{a}$'s and $\mathtt{b}$'s.
Note that the use of $\mathtt{cons}$ and 
$\mathtt{nil}$ could be regarded as overloading, but this is not
forbidden by~\cite{LS01} as it is not in contradiction to the grammar
being deterministic. 

\setlength{\unitlength}{3mm}
\begin{figwindow}[0,r,%
{\mbox{\begin{picture}(11,4.5)%
{\linethickness{0.2\unitlength}
\put(1,0){\framebox(2,3){$LL$}}}%
\curve(3,2, 4,2.5, 4,4, 2.5,4, 2,3)%
\put(2,3){\vector(0,-1){0}}%
\put(3,1){\vector(1,0){2}}%
\put(5,0){\framebox(2,3){$L$}}%
\curve(7,2, 8,2.5, 8,4, 6.5,4, 6,3)%
\put(6,3){\vector(0,-1){0}}%
\put(7,1){\vector(1,0){2}}%
\put(9,0){\framebox(2,3){$E$}}%
\end{picture}}},%
{List of lists \label{listlist-type-graph-fig}}]
We use the usual list notation for ease of reading.
The type graph of $LL$ is shown in
Fig.~\ref{listlist-type-graph-fig}. 
We are interested in the labelling of all non-terminals reachable from 
$LL$. 
Let $t = [[\mathtt{a}],[\mathtt{b}]]$. Then 
$\zeta(LL,E,t) = \zeta(LL,L,t) = \zeta(LL,LL,t) = \emptyset$. 
Now let $t = [[\mathtt{a}],[\mathtt{X}]]$. Then 
$\zeta(LL,E,t) = \set{\mathtt{X}}$ and
$\zeta(LL,L,t) = \zeta(LL,LL,t) = \emptyset$.
Now let $t = [[\mathtt{a}],\mathtt{X}]$. Then
$\zeta(LL,E,t) = \emptyset$, 
$\zeta(LL,L,t) = \set{\mathtt{X}}$ and
$\zeta(LL,LL,t) = \emptyset$.

\end{figwindow}
\end{ex}

\subsection{Labelling and Polymorphism}\label{labelling-polymorphic-subsec}
We now want to adapt the idea of labelling to the case when we have
polymorphism. With polymorphism, one can have infinitely many types,
and even though the type graph for a fixed type as starting node is
finite, it can become arbitrarily large, i.e., it can have an
arbitrary number of non-terminals reachable from the starting node. 
Also, it would clearly be desirable to describe the labellings for say
$\mathtt{list(int)}, \mathtt{list(list(int))},\dots$ in a
uniform way. This motivates defining a hierarchy in the type
graph. 

\begin{defi} \label{typerelations-def}
A type $\sigma$ is a {\bf recursive type of $\phi$} 
(denoted as $\sigma \rec \phi$) if 
$\sigma \sub \phi$ and $\phi \sub \sigma$.
%If $\{ \sigma_1,\dots,\sigma_m \} = 
%\{ \sigma \mid \sigma \rec \phi,\ \sigma\neq\phi \}$, 
%and 
%$\sigma_j \prec \sigma_{j+1}$ 
%for all $j \in \{1, \dots, m-1 \}$, 
%we write $\Rec(\phi)$ for the tuple
%$\langle  \sigma_1,\dots,\sigma_m \rangle$.
We write $\Rec(\phi)$ for the tuple of recursive types of $\phi$ 
other than $\phi$ itself, ordered by $\prec$ (see Sec.~\ref{prelim-sec}).

A type
$\sigma$ is a {\bf non-recursive subterm type (NRS) of $\phi$} 
(denoted as $\sigma \nrs \phi$)
if $\phi \notsub \sigma$ and there is
a type $\tau$ such that 
$\sigma \subone \tau$ and $\tau \rec \phi$. 
%If $\{ \sigma_1,\dots,\sigma_m \} = 
%\{ \sigma \mid \sigma \nrs \phi \}$, 
%and 
%$\sigma_j \prec \sigma_{j+1}$ 
%for all $j \in \{1, \dots, m-1 \}$, 
%we write $\Nrs(\phi)$ for the tuple
%$\langle  \sigma_1,\dots,\sigma_m \rangle$.
We write $\Nrs(\phi)$ for the tuple of non-recursive subterm types of
$\phi$, 
ordered by $\prec$.
\end{defi}

It follows immediately from the definition that, for any types $\phi,\sigma$, 
we have $\phi \rec \phi$ and,
if $\sigma \nrs \phi$, then $\sigma \not\rec \phi$.
Consider the type graph for $\phi$.
The recursive types of $\phi$ are all the types in the
strongly connected component (SCC) containing $\phi$.
The non-recursive
subterm types of $\phi$ are all the types $\sigma$ not in the SCC but
such that there is an edge from the SCC containing $\phi$ to $\sigma$.

\begin{ex}\label{typerelations-ex}
Consider Fig.~\ref{type-graphs-fig}.
Let
$\mathcal{F}_\mathtt{LISTS} = \set{\mathtt{nil_{\rar list(U)}, 
                            cons_{U,list(U) \rar list(U)}}}$.
We have
$\mathtt{list(U)} \rec \mathtt{list(U)}$ and
$\mathtt{U} \nrs \mathtt{list(U)}$. 

Let
$\mathcal{F}_\mathtt{NESTS} = \mathcal{F}_\mathtt{LISTS} \cup \set{\mathtt{
                            e_{V \rar nest(V)},
                            n_{list(nest(V)) \rar nest(V)}}}$. 
The \texttt{NESTS} language implements {\em rose trees}~\cite{M88},
i.e., trees where the number of children of each node is not fixed. 
We have
$\mathtt{list(nest(V))} \rec \mathtt{nest(V)}$ and
$\mathtt{nest(V)} \rec \mathtt{nest(V)}$ and
$\mathtt{V} \nrs \mathtt{nest(V)}$. 

Suppose $\mathcal{F}_\mathtt{STRINGS}$ contains all strings with 
$\rar\mathtt{str}$ as type subscript.
Let 
$\mathcal{F}_\mathtt{TABLES} = \mathcal{F}_\mathtt{STRINGS} \cup$
\[
\{\mathtt{
lh_{\rar bal}, rh_{\rar bal}, eq_{\rar bal}, null_{\rar table(U)}},
\mathtt{node_{table(U),str,U,bal,table(U) \rar table(U)}}\},
\] 
 The type $\mathtt{bal}$ contains three constants
 representing balancing information. We have 
$\mathtt{table(U)} \rec \mathtt{table(U)}$ and
$\Nrs(\mathtt{table(U)}) = 
\langle \mathtt{U}, \mathtt{bal}, \mathtt{str} \rangle$.

An NRS of a flat type is often just a parameter of that type, as in 
$\mathtt{U} \nrs \mathtt{list(U)}$. 
However, this is not always the case, as witnessed by
$\mathtt{str} \nrs \mathtt{table(U)}$.
\end{ex}

Instead of looking at the labellings of \emph{all} non-terminals
reachable from some starting node without distinction~\cite{LS01}, we classify
them according to the recursive types and the NRSs of that node. This
will be reflected in the construction of abstract domains.

Definition~\ref{typerelations-def} is obviously applicable in particular in the
monomorphic case and thus to the grammars as in~\cite{LS01}.
Figure~\ref{listlist-type-graph-fig} shows that the recursive types
and NRSs may not be all types reachable from a starting node. In
that example, we have $LL\rec LL$ and  $L\nrs LL$. 
In the approach of~\cite{LS01}, we may also be
interested in $\zeta(LL,E,t)$ for some term $t$, so in the labellings
of $E$. In the approach proposed here, the domain construction for
$LL$ depends on $E$ only indirectly, via the abstract domain for $L$. 
Without such an inductive approach to domain construction, we would
not know how to deal with polymorphism. 

The key to devising a ``parametric'' abstract domain construction is
to focus on type constructors, or equivalently, on flat types
$c(\vect{u})$. So for example, we should focus on $\mathtt{list(U)}$
and not a particular instance such as $\mathtt{list(int)}$. This
may not be surprising, but it has two consequences which may not be
obvious.

First, note that the relation $\nrs$ is not stable under instantiation
of types. This can be seen by comparing \texttt{LISTS} with
\texttt{NESTS}. We have $\mathtt{U} \nrs \mathtt{list(U)}$, but
$\mathtt{nest(V)} \rec \mathtt{list(nest(V))}$. The abstract
domain for $\mathtt{list(nest(V))}$ however, being derived from 
the abstract domain for $\mathtt{list(U)}$, must relate to
$\mathtt{nest(V)}$ as if $\mathtt{nest(V)}$ was an NRS of 
$\mathtt{list(nest(V))}$. In contrast, the abstract domain for
$\mathtt{nest(V)}$ must reflect that 
$\mathtt{list(nest(V))} \rec \mathtt{nest(V)}$. 
One could paraphrase this by saying: 
\texttt{LISTS} does not know about \texttt{NESTS}, but 
\texttt{NESTS} does know about \texttt{LISTS}. 

The second consequence can be illustrated using \texttt{TABLES}. The type
$\mathtt{table(U)}$ has three NRSs, and each of them will be dealt 
with in the construction of the abstract domain. However, the instance 
$\mathtt{table(string)}$ has only two NRSs, as $\mathtt{U}$
becomes instantiated to $\mathtt{string}$. The
domain for $\mathtt{table(\tau)}$ will be based on assuming 
{\em three} NRSs, i.e., it will deal with the value and the key
arguments separately, even if by coincidence $\tau$ is equal to
$\mathtt{string}$.  

We now define a function $\mathcal{Z}$ in analogy to $\zeta$.  In the
approach of~\cite{LS01}, we could safely identify a grammar with its
starting non-terminal. In what follows, we will always assume a
grammar $\mathcal{G}(\phi)$ where $\phi$ is flat 
(Def.~\ref{corresponding-grammar-def}). However, it is also 
useful to consider productions of that grammar starting from
some other non-terminal than the ``official'' starting non-terminal.
Therefore
$\mathcal{Z}$ has {\em four} arguments, the additional first one
specifying the grammar and the second the starting symbol. 

Unlike $\zeta$, the function $\mathcal{Z}$ also collects non-variable terms. 

\begin{defi}\label{my-labelling-def}
Let $\phi$ be a flat type, $\tau$ be a type such that 
$\tau\rec\phi$, and $\sigma$ a type such that either 
$\sigma\rec\phi$ or $\sigma\nrs\phi$. 

We denote by $\mathcal{Z}(\phi,\tau,\sigma,t)$ the function
which returns the set of all terms $s$ such that 
$\nt{\tau}(t) \rar^* \nt{\sigma}(s)$ in the grammar $\mathcal{G}(\phi)$.
\end{defi}

The function $\mathcal{Z}$ is lifted to sets (in the fourth argument) in the obvious way.

\begin{ex}\label{my-labelling-ex}
Let 
$\mathcal{F} = \mathcal{F}_\mathtt{LISTS} \cup 
\set{\mathtt{a_{char}}, \mathtt{b_{char}}}$. 
We have
\begin{eqnarray*}
\mathcal{Z}(\mathtt{list(U)}, \mathtt{list(U)}, \mathtt{list(U)}, [\mathtt{a},\mathtt{X}]) & = &
\set{[\mathtt{a},\mathtt{X}], [\mathtt{X}], []}\\
\mathcal{Z}(\mathtt{list(U)}, \mathtt{list(U)}, \mathtt{U},
[\mathtt{a},\mathtt{X}]) & = & 
\set{\mathtt{a, X}}\\[1ex]
\mathcal{Z}(\mathtt{list(U)}, \mathtt{list(U)},\mathtt{list(U)}, [[\mathtt{a}],[\mathtt{X}]]) & = & 
\set{[[\mathtt{a}], [\mathtt{X}]], [[\mathtt{X}]], []}\\ 
\mathcal{Z}(\mathtt{list(U)}, \mathtt{list(U)},\mathtt{U},[[\mathtt{a}], [\mathtt{X}]]) & = & 
\set{[\mathtt{a}], [\mathtt{X}]}\\[1ex]  
\mathcal{Z}(\mathtt{list(U)}, \mathtt{list(U)}, \mathtt{list(U)},\mathtt{[[a]|X]}) & = & 
\set{\mathtt{[[a]|X]}, \mathtt{X}}\\
\mathcal{Z}(\mathtt{list(U)}, \mathtt{list(U)}, \mathtt{U},\mathtt{[[a]|X]}) & = & 
\set{[\mathtt{a}]}.  
\end{eqnarray*}
Note that unlike $\zeta$ (Ex.~\ref{labelling-ex}), 
$\mathcal{Z}$ cannot be used to 
extract from the term $[[\mathtt{a}],[\mathtt{X}]]$ the subterm
$\mathtt{X}$ directly.  

Now consider the \texttt{NESTS} example, assuming that
$\mathcal{F}_\mathtt{NESTS}$ is augmented with the integers
$1_\mathtt{int},\dots$. We have
\begin{eqnarray}
\mathcal{Z}(\mathtt{nest(V), nest(V), nest(V), n([e(7)])}) & = &
\set{\mathtt{n([e(7)]), e(7)}}\nonumber\\
\mathcal{Z}(\mathtt{nest(V), nest(V), list(nest(V)), n([e(7)])}) & = &
\set{\mathtt{[e(7)], []}}\nonumber\\
\mathcal{Z}(\mathtt{nest(V), nest(V), V, n([e(7)])}) & = &
\set{\mathtt{7}}\nonumber\\[1ex]
\mathcal{Z}(\mathtt{nest(V), list(nest(V)), nest(V), [n([e(7)])]}) & = &
\set{\mathtt{n([e(7)]), e(7)}}\label{list-as-nest-1}\\
\mathcal{Z}(\mathtt{nest(V), list(nest(V)), list(nest(V)), [n([e(7)])] }) & = &
\set{\mathtt{[n([e(7)])], [e(7)], []}}\label{list-as-nest-2}\\
\mathcal{Z}(\mathtt{nest(V), list(nest(V)), V, [n([e(7)])]}) & = &
\set{\mathtt{7}}\label{list-as-nest-3}\\[1ex]
\mathcal{Z}(\mathtt{list(U), list(U), list(U), [n([e(7)])]}) & = &
\set{\mathtt{[n([e(7)])], []}}\label{list-as-list-1}\\
\mathcal{Z}(\mathtt{list(U), list(U), U, [n([e(7)])]}) & = &
\set{\mathtt{n([e(7)])}}\label{list-as-list-2}\\[1ex]
\mathcal{Z}(\mathtt{nest(V), nest(V), V, e(7)}) & = &
\set{\mathtt{7}}\label{nest-plain}
\end{eqnarray}
Note the difference between the labellings obtained for 
$\mathtt{[e(7)]}$ depending on whether we use the grammar for 
$\mathtt{nest(V)}$ (\ref{list-as-nest-1}, \ref{list-as-nest-2}, \ref{list-as-nest-3}), or
the grammar for $\mathtt{list(U)}$ (\ref{list-as-list-1}, \ref{list-as-list-2}).
\end{ex}

%PATCH 1

\begin{rmlemma}\label{subterm-label-lem}
Let $\phi$ and $\tau$ be flat types such that 
$\appl{\Theta}{\tau} \rec \phi$ for some  $\Theta$.
Let $t = \ftau(t_1,\dots,t_n)$ be a term such that
$\_\vdash t:\tau\Theta\Theta'$ for some $\Theta'$.
Then
\[
\mathcal{Z}(\phi,\tau\Theta,\sigma,t) =
\left\{
\begin{array}{llll}
& \set{t_i \mid \tau_i\Theta = \sigma} \cup & 
\displaystyle
\bigcup_{\tau_i\Theta\rec\phi} \mathcal{Z}(\phi,\tau_i\Theta,\sigma,t_i) \ &
\mbox{if}\ \sigma\nrs\phi\\
 & & 
\displaystyle
\bigcup_{\tau_i\Theta\rec\phi} \mathcal{Z}(\phi,\tau_i\Theta,\sigma,t_i) &
\mbox{if}\ \sigma\rec\phi, \sigma\neq\phi\\
\set{t} \cup & & 
\displaystyle
\bigcup_{\tau_i\Theta\rec\phi} \mathcal{Z}(\phi,\tau_i\Theta,\sigma,t_i) &
\mbox{if}\ \sigma=\phi.
\end{array}
\right.
\]
\end{rmlemma}

\begin{pf}
Follows from the fact that for each $i\in\oneton$, we have
$\nt{\tau\Theta}(t) \rar \nt{\tau_i\Theta}(t_i)$ in 
$\mathcal{G}(\phi)$. 
\end{pf}

\section{Abstract Terms}\label{abstract-terms-sec}
In this section, we define an {\em abstraction} of terms using the
notions of recursive type and non-recursive subterm type. This amounts
to a generalisation of~\cite{CL00}.

\subsection{Abstract Domains and Terms}\label{abstract-terms-subsec}
We first introduce the formalism of {\em set logic programs} shown to be
powerful for program analyses~\cite{LS01}. Consider a language
based on a set of variables $\mathcal{V}$ and a set of functions
$\mathcal{F}^\oplus = \set{\emptyset,\oplus}$, where
$\emptyset/0$ represents the empty set and 
$\oplus/2$ is a set constructor. {\em Set expressions} are elements of
the
term algebra $\mathcal{T}(\mathcal{F}^\oplus,\mathcal{V})$ modulo the
ACI1 equality theory, consisting of:
\begin{equation}
\label{aci1-eq}
\begin{array}{rcllrcll}
(x\oplus y) \oplus z & \!=\! & x \oplus (y\oplus z) \!\! &
\mbox{(associativity)} &
x\oplus x & \!= \!& x &
\mbox{(idempotence)} \\
x \oplus y & \!=\! & y \oplus x &
\mbox{(commutativity)}  &
x \oplus \emptyset & \!=\! & x &
\mbox{(unity)}
\end{array}
\end{equation}
Lagoon and Stuckey~\cite{LS01} now proceed by regarding each
non-terminal in the grammar corresponding to a variable $x$ as
an {\em abstract variable}. The instantiation of that abstract
variable obtained from the execution of the abstract program gives us
information about the labels of the non-terminal after the concrete
execution. 

%\comment{Maybe those three paragraphs can be shortened, noting that 
%what we do very much follows~\cite{CL00}.}
%Lagoon and Stuckey now proceed as follows. First, it must be noted that
%each program variable $X$ is associated with a grammar
%$\mathcal{G}_X$, whose non-terminals are uniquely marked as belonging
%to $X$. In the abstraction, each non-terminal in $\mathcal{G}_X$ 
%becomes an ``abstract variable''. For example, assuming that $X$ has
%the grammar in Ex.~\ref{labelling-ex} associated with it, then if the
%execution of the abstract program results in $LL = \emptyset$, this
%means that $\zeta(LL,LL,X\theta) = \emptyset$, where $\theta$ is the
%substitution obtained from the concrete execution.

%We prefer to define the abstraction in a slightly different
%way. First, we do not introduce new variables; the variables of the
%abstraction of a clause are the variables of the original clause.  
%The essential reason for this is that in the presence of polymorphism, 
%any polymorphic instance of a clause can be used in a program
%execution, and the number of abstract variables we would have to 
%introduce would somehow have to depend on that instance. 

We do not see how this approach could work in the presence of
polymorphism. Instead, we follow~\cite{CL00}.
We now introduce new function symbols, one $c^\mathcal{A}$ for
each type constructor $c\in\mathcal{K}$, in addition to $\emptyset$
and $\oplus$. These are used to collect the information corresponding
to the different non-terminals in a structure, which we call 
{\em abstract term}. The arity of $c^\mathcal{A}$ is given by the
arity of $\Nrs(c(\bar{u}))$ plus the arity of $\Rec(c(\bar{u}))$.

\begin{defi}\label{domain-def}\label{alpha-def}
We define
\[
\mathcal{F}^\mathcal{A} := 
\mathcal{F}^\oplus \cup
\set{c^\mathcal{A}/m \mid c\in\mathcal{K},\ 
m = \#(\Nrs(c(\bar{u}))) + \#(\Rec(c(\bar{u})))}.
\]
Now let $\tau = c(\bar{u})$, 
$\Nrs(\tau) = \langle \rho_1,\dots,\rho_{m'} \rangle$, and
$\Rec(\tau) = \langle \rho_{m'+1},\dots,\rho_m \rangle$.
For a term 
$t = \ftau(t_1,\dots,t_n)$, we define
\[
\alpha(t) = 
c^\mathcal{A}
\left(
\bigoplus_{\tau_i=\rho_1} \alpha(t_i),
\dots,
\bigoplus_{\tau_i=\rho_m} \alpha(t_i)
\right)
\oplus
\bigoplus_{\tau_i=\tau} \alpha(t_i).
\]
For a variable $x$ we define $\alpha(x) = x$.
We call the image of $\alpha$ the {\bf domain of abstract terms}, or
simply the {\bf abstract domain}.
\end{defi}

In~\cite{CL96}, the abstraction function is denoted $\mathit{type}$,
and it is essentially a special case of the above definition for 
$\#(\Nrs(c(\bar{u})))  = 1$ and $\#(\Rec(c(\bar{u}))) = 0$.
Using our terminology, they assume that there is a
function, which they denote by $\alpha$, returning for each 
$f_{\dots\rar c(u)}$ the type constructor $c$; moreover, there is a
function $\pi$ returning the set of argument positions of $f$ that have
declared type $c(u)$, and all other argument positions are assumed to
have declared type $u$. However, since their typing is descriptive,
$\alpha$ and $\pi$ have to be provided by the user.

\begin{ex}\label{alpha-ex}
Consider again Ex.~\ref{my-labelling-ex}.
\[
\begin{array}{rclcl}
\alpha(\mathtt{7}) & = &
\mathtt{int}^\mathcal{A}\\
\alpha(\mathtt{[7]}) & = &
\mathtt{list^\mathcal{A}(\alpha(7)) \oplus \alpha(nil)} & = & 
\mathtt{list^\mathcal{A}(int^\mathcal{A}) \oplus list^\mathcal{A}(\emptyset)}\\
\alpha(\mathtt{e(7)}) & = &
\mathtt{nest^\mathcal{A}(\alpha(7),\emptyset)} & = & 
\mathtt{nest^\mathcal{A}(int^\mathcal{A},\emptyset)}\\
\alpha(\mathtt{n([e(7)])}) & = &
\mathtt{nest^\mathcal{A}(\emptyset,\alpha([e(7)]))} & = & 
\mathtt{nest^\mathcal{A}(\emptyset,list^\mathcal{A}(nest^\mathcal{A}(int^\mathcal{A},\emptyset)))}.
\end{array}
\]
\end{ex}

Note how it comes into play that the empty $\bigoplus$-sequence is
naturally defined as  the neutral element $\emptyset$. There
is a notable difference between~\cite{CL96} and~\cite{CL00} at this point.
In the latter, there is no neutral element. This means
in particular that $\mathtt{nil}$ cannot be abstracted as 
$\mathtt{list(\emptyset)}$. Instead, it is abstracted as
$\mathtt{nil}$, and as a consequence, the list $\mathtt{[7]}$ is abstracted as 
$\mathtt{list(int) \oplus nil}$. While it is argued that such an
abstraction simplifies the implementation of abstract unification, we believe that an
object $\mathtt{list(int) \oplus nil}$ mixes types/abstract terms on
the one hand and concrete terms on the other hand in a way which is 
undesirable.

In fact, from the design of our abstract domains and the fact that we
are analysing prescriptively typed programs, it follows that 
whenever an expression
$c^\mathcal{A}(\dots) \oplus {c'}^\mathcal{A}(\dots)$ occurs, then
$c=c'$. This also explains why in the definition of $\alpha$, the
abstraction of those $t_i$ such that $\tau_i\rec\tau$ but 
$\tau_i\neq\tau$ is included in reserved argument positions of 
$c^\mathcal{A}(\dots)$, whereas the abstraction of those $t_i$ 
such that $\tau_i=\tau$ is directly conjoined (using $\oplus$) with
the whole expression $c^\mathcal{A}(\dots)$.

Looking at Ex.~\ref{alpha-ex}, one might expect that 
$\mathtt{list^\mathcal{A}(int^\mathcal{A}) \oplus list^\mathcal{A}(\emptyset)}$
can be simplified to
$\mathtt{list^\mathcal{A}(int^\mathcal{A})}$. Maybe less obvious, one
might also expect that the abstract term
$
\mathtt{nest^\mathcal{A}(\emptyset,list^\mathcal{A}(nest^\mathcal{A}(int^\mathcal{A},\emptyset)))}
$
can be simplified to 
$\mathtt{nest^\mathcal{A}(int^\mathcal{A},\emptyset)}$. We now extend
ACI1 by further axioms for this purpose.

\begin{defi}\label{axioms-def}
For each 
$c^\mathcal{A}/m \in \mathcal{F}^\mathcal{A}$, the  
{\bf distributivity axiom} is defined as follows:
\begin{equation}
\label{distr-eq}
c^\mathcal{A} (x_1,\dots,x_m) \oplus
c^\mathcal{A} (y_1,\dots,y_m) = 
c^\mathcal{A} (x_1 \oplus y_1,\dots, x_m \oplus y_m) 
\end{equation}
Moreover, consider a flat type 
$\phi=d(\vect{v})$ such that 
$\Nrs(\phi) = \langle \sigma_1,\dots,\sigma_{l'}\rangle$,
$\Rec(\phi) = \langle \sigma_{l'+1},\dots,\sigma_{l}\rangle$.
For each $j\in\set{{l'}+1,\dots,l}$, we have
$\sigma_j = \tau\Theta$ for some flat type 
$\tau = c(\vect{u})$ and some $\Theta$.  Suppose
$\Nrs(\tau) = \langle \rho_1,\dots,\rho_{m'}\rangle$,
$\Rec(\tau) = \langle \rho_{m'+1},\dots,\rho_{m}\rangle$,
We define the 
{\bf extraction axiom} for $d^\mathcal{A}$ and $\sigma_j$ as follows:
\[
\label{extraction-eq}
\begin{array}{l}
d^\mathcal{A}(x_1,\dots,x_{j-1},
              c^\mathcal{A}(y_1,\dots,y_m)\oplus x_j,
              x_{j+1},\dots,x_l) = \\ 
\displaystyle
d^\mathcal{A}\!\left(\!
              x_1\oplus\!\!\!\bigoplus_{\rho_k\Theta = \sigma_1}\!\!\! y_k,
              \;\dots,\;
              x_{j-1}\oplus\!\!\!\!\!\bigoplus_{\rho_k\Theta = \sigma_{j-1}}\!\!\!\!\! y_k,
              \;x_j,\;
              x_{j+1}\oplus\!\!\!\!\!\bigoplus_{\rho_k\Theta = \sigma_{j+1}}\!\!\!\!\! y_k,
              \;\dots,\;
              x_m\oplus\!\!\!\bigoplus_{\rho_k\Theta = \sigma_l}\!\!\! y_k
              \! \right) \\
\multicolumn{1}{r}{\displaystyle \oplus
              \bigoplus_{\rho_k\Theta = \phi} y_k.}
\end{array}
\]
Let {\bf ACI1DE} be the theory given by the axioms in~(\ref{aci1-eq}) and
the distributivity and extraction axioms. We abbreviate ACI1DE by 
{\bf \theory} and denote equality modulo \theory\ as $=_\mathrm{\theory}$.
\end{defi}

Note that applying a distributivity or extraction axiom from left to
right decreases the number of occurrences of function symbols by 1.

\begin{ex}\label{extractor-ex}
Consider \texttt{LISTS}, \texttt{NESTS}, respectively. 
The extraction axiom for $\mathtt{nest(V)}$ and $\mathtt{list(nest(V))}$
is
\[
\mathtt{nest}^\mathcal{A}(x_1,\mathtt{list}^\mathcal{A}(y)\oplus x_2)
= 
\mathtt{nest}^\mathcal{A}(x_1, x_2) \oplus y.
\]
In \theory, we have
\[
\begin{array}{l}
\mathtt{list^\mathcal{A}(int^\mathcal{A}) \oplus list^\mathcal{A}(\emptyset)}
=_\mathrm{\theory} 
\mathtt{list^\mathcal{A}(int^\mathcal{A}\oplus \emptyset)} 
=_\mathrm{\theory} 
\mathtt{list^\mathcal{A}(int^\mathcal{A})} \\ 
\mathtt{nest^\mathcal{A}(\emptyset,list^\mathcal{A}(nest^\mathcal{A}(int^\mathcal{A},\emptyset)))} 
=_\mathrm{\theory} 
\mathtt{nest^\mathcal{A}(\emptyset,\emptyset)\oplus nest^\mathcal{A}(int^\mathcal{A},\emptyset)}
=_\mathrm{\theory} \\ 
\multicolumn{1}{r}{
\mathtt{nest^\mathcal{A}(int^\mathcal{A},\emptyset)}.}
\end{array}
\]
\end{ex}

In~\cite{CL00}, it is mentioned that one might have chosen to add a
distributivity axiom but the authors argue the case for not doing
so. The extraction axiom has no equivalent in~\cite{CL00}.

\subsection{Normal Abstract Terms}
\label{normal-abstract-terms-subsec}
Following~\cite{CL00}, we have defined the abstraction by structural
induction on a term. This definition is a good basis for our analysis,
but it is still unsatisfactory: as it is stated (i.e.~without
applying any axioms), even for ground terms, the abstraction of a term
is proportional in size to the term itself; consequently, the
abstraction is not in a form that makes it convenient to read any
properties of the concrete term from it.

We first show that using \theory, any abstract term can be converted in
a normal form. To this end, it is useful to view abstract terms as
typed terms according to the rules of Table~\ref{rules-tab}. 
For each $\tau = c(\vect{u})$ with 
$\Nrs(\tau) = \langle \rho_1,\dots,\rho_{m'} \rangle$ and
$\Rec(\tau) = \langle \rho_{m'+1},\dots,\rho_m \rangle$, we 
declare 
$c^\mathcal{A}_{\rho_1,\dots,\rho_m \rar \tau}$. 
Moreover, we declare $\emptyset_{\rar u}$ and
$\oplus_{u,u\rar u}$.\footnote{These declarations violate the Simple
Range Condition, and in any case would not be permissible in any
existing typed programming language since a range type must not be a 
parameter. However, this causes no problems for our theoretical 
considerations.}
Those declarations are designed exactly so that the following
proposition holds.

\begin{propo}\label{abstract-nrs-rec-prop}
The $\nrs$- and $\rec$-relations based on the declared types of 
the ``abstract functions'' in $\mathcal{F^A}$ are the same as 
the  $\nrs$- and $\rec$-relations based on $\mathcal{F}$.
\end{propo}

To distinguish type judgements in the concrete
and abstract language, we use $\vdash^\mathcal{A}$ for the latter.
The following proposition says that the abstraction of a term
has the same type as the concrete term itself. Its proof is
straightforward by structural induction.

\begin{propo}\label{abstract-term-is-typed-prop}
If $\Gamma\vdash t:\tau$ then $\Gamma\vdash^\mathcal{A} \alpha(t) : \tau$.
\end{propo}

The next lemma says that application of the equality axioms
preserves well-typing. The interesting axiom is the extraction axiom.

\begin{rmlemma}\label{axioms-preserve-well-typing-lem}
If $\Gamma\vdash^\mathcal{A} a:\tilde{\tau}$ and $a =_\mathrm{\theory} b$ then
$\Gamma\vdash^\mathcal{A} b:\tilde{\tau}$.
\end{rmlemma}

\begin{pf}
We only consider the extraction axiom. Assume the notations of
Def.~\ref{axioms-def}, and consider an abstract term 
\[
a =  d^\mathcal{A}(b_1,\dots,b_{j-1},
              c^\mathcal{A}(a_1,\dots,a_m)\oplus b_j,
              b_{j+1},\dots,b_l)
\] 
where
$ \_ \vdash^\mathcal{A} a :\phi\Theta'$
for some $\Theta'$ (note that the typing rules are such that $a$
 must have a type that is an instance of $\phi$, but not
necessarily $\phi$ itself). 
By the rules of Table~\ref{rules-tab}, we must have
$\_ \vdash^\mathcal{A} b_{j'}: \sigma_{j'}\Theta'$ for all $j'\in\onetol$ and
$\_ \vdash^\mathcal{A} a_k: \rho_k\Theta\Theta'$ for all $k\in\onetom$.
Therefore it follows that for $j'\in\onetol\setminus\set{j}$, we have
\[
\_\vdash^\mathcal{A} 
\left(
b_{j'}\oplus\bigoplus_{\rho_k\Theta = \sigma_{j'}} a_k
\right) : \sigma_{j'}\Theta',\ 
\mbox{
and we also have}\ 
\_\vdash^\mathcal{A} 
\left(
\bigoplus_{\rho_k\Theta = \phi} a_k
\right) : \phi\Theta'.
\]
This implies that if $b$ is obtained from $a$ by applying the
extraction axiom in the $j$th position, then 
$ \_ \vdash^\mathcal{A} b :\phi\Theta'$.
\end{pf}

We now define normal forms for abstract terms. 
To simplify the
notation, we denote a variable sequence $x_1\oplus\dots\oplus x_n$ as 
$x^\oplus$, and of course, this is $\emptyset$ if $n=0$.
Note that the following definition is by structural induction: a
normal abstract term for $\tau\Theta$ is defined based on a normal
abstract term for some $\tau_i\Theta$. The well-foundedness of this
induction is not obvious but has been stated in~\cite[Lemma 4.3]{SHK00}.

\begin{defi}\label{normal-form-def}
For a parameter, the only {\bf normal abstract term} is $\emptyset$.

Now let $\tau=c(\vect{u})$ be a flat type such that 
$\Nrs(\tau) = \langle\rho_{1},\dots,\rho_{m'}\rangle$ and
$\Rec(\tau) = \langle \rho_{m'+1},\dots,\rho_m \rangle$, and $\Theta$ be
any type substitution. A {\bf normal abstract term for $\tau\Theta$}
is $\emptyset$ or of the form
$c^\mathcal{A}(a_1 \oplus x^\oplus_1 ,\dots,
               a_{m'} \oplus x^\oplus_{m'}, 
               x^\oplus_{m'+1},\dots,x^\oplus_m) \oplus x^\oplus$, where
for each $i\in\set{1,\dots,m'}$, $a_i$ is a normal abstract term for
$\rho_i\Theta$. 
\end{defi}

Note that in a normal abstract term, the ``second half'' of argument
positions, i.e., those corresponding to recursive subterm types of
$\tau$ other than $\tau$ itself, must only contain
variables. Considering again Def.~\ref{alpha-def}, intuitively the
abstractions of the recursive subterms of $t$ are stored in those
positions only temporarily. The ultimate goal is to remove any
non-variables there using the axioms, in particular the extraction
axiom. 

Based on the propositions above, one can show:

\begin{rmtheorem}\label{normal-form-exists-thm}
For any $t$ with $\_\vdash t:\phi$, $\alpha(t)$ has a representative 
which is a normal abstract term for $\phi$.
\end{rmtheorem}

\begin{pf}
By Prop.~\ref{abstract-term-is-typed-prop}
and Lemma~\ref{axioms-preserve-well-typing-lem} we have
$\_\vdash a:\phi$ if $a=_\mathrm{\theory} \alpha(t)$.

Assume the notations of Def.~\ref{axioms-def}.
The fact that an abstract term is typed according to the rules of
Table~\ref{rules-tab} means in particular that if it
has the form
$d^\mathcal{A}(b_1,\dots,b_{j-1},
              \dots\oplus {c'}^\mathcal{A}(\dots)\oplus\dots,
              b_{j+1},\dots,b_l)$ where $j\in\set{l'+1,\dots,l}$,
i.e., it is not in normal form, then $c' = c$ and an extraction 
axiom is applicable, possibly after several applications of
associativity and commutativity.

Likewise, if an abstract term has the form
$\dots d^\mathcal{A}(\dots) \oplus\dots\oplus 
{d'}^\mathcal{A}(\dots) \dots$, then $d = d'$ and the distributivity
axiom is applicable.

Since the distributivity and extraction axioms can only be applied a
finite number of times, it follows that successive application of them 
yields an abstract term in normal form.
\end{pf}

Example~\ref{extractor-ex} shows the conversion of two 
abstract terms to their normal forms.

We can make some further observations. The first follows
from the definition of $\alpha$.

\begin{propo}\label{alpha-variables-prop}
For a term $t$, $\alpha(t)$ contains variables if and only if $t$
contains variables.
\end{propo}

The next proposition follows from the fact that by the transparency
condition, no subterm type of a ground type $\phi$ can contain a
parameter.

\begin{propo}\label{exactly-one-normal-prop}
If $\phi$ is a ground type, then there is exactly one normal abstract
term for $\phi$ not containing variables and not containing
$\emptyset$.

Let $a$ be this abstract term. Then any other normal abstract term for $\phi$
not containing variables is obtained by replacing some subterms in $a$
with $\emptyset$.
\end{propo}

The previous three statements tell us that the size of the abstraction
of a ground term depends only on its type and not on the size of the
term itself. However, it would be wrong to conclude that all ground
terms of the same type have the same abstraction. The problem is due
to polymorphism: one term can have several types. For example, it is
correct to say that both $\mathtt{[]}$ and $\mathtt{[7]}$ are of 
type $\mathtt{list(int)}$, but 
$\alpha([]) = \mathtt{list^\mathcal{A}(\emptyset)}$ and 
$\alpha(\mathtt{[7]}) = \mathtt{list^\mathcal{A}(int^\mathcal{A})}$.
However, we can state:

\begin{rmlemma}\label{same-type-same-alpha-lem}
If $\vdash s:\phi$ and $\vdash t:\phi$ and $s,t$ are ground, then
both $\alpha(s)$ and $\alpha(t)$ are obtained from the unique normal
abstract term as mentioned in Prop.~\ref{exactly-one-normal-prop} by
replacing zero or more subterms with $\emptyset$.
\end{rmlemma}

\section{Relating the Abstraction and the Labels}
\label{alpha-label-relation-sec}
Now that we know that each abstract term has a representative in a
compact normal form, we state a theorem which relates the abstraction
of a term to the labels as defined in Sec.~\ref{structure-sec}. 
Actually, it would have been possible to have the following theorem as 
{\em definition} of $\alpha$, and have our present definition as a
lemma. This is effectively what we did in~\cite{SHK00}. 
The theorem also links~\cite{CL00} with~\cite{LS01}.
Note that $\alpha$ is lifted to sets as follows:
$\alpha(S) := \bigoplus_{t\in S} \alpha(t)$. 

Before we can show the theorem, we extend the definition of
$\mathcal{Z}$ to abstract terms, typed as shown in the previous
section. We state the following lemma.

\begin{rmlemma}\label{abst-conc-labelling-lem}
Let $\phi$ and $\tau$ be flat types such that 
$\appl{\Theta}{\tau} \rec \phi$ for some  $\Theta$. Let $t$ be a term
such that
$\_\vdash t:\tau\Theta\Theta'$ for some $\Theta'$.
Then for any $\sigma\nrs\phi$, or $\sigma\rec\phi$, we have
$\alpha(\mathcal{Z}(\phi,\tau\Theta,\sigma,t)) =
\mathcal{Z}(\phi,\tau\Theta,\sigma,\alpha(t))$.
\end{rmlemma} 

\begin{pf}
The proof is by induction on the depth of $t$. 
First suppose that $t\in\mathcal{V}$. Then we have to distinguish
whether $\sigma = \tau\Theta$ or $\sigma\neq\tau\Theta$. In the first case,
$\alpha(\mathcal{Z}(\phi,\tau\Theta,\sigma,t)) = t =
\mathcal{Z}(\phi,\tau\Theta,\sigma,\alpha(t))$. In the second case,
$\alpha(\mathcal{Z}(\phi,\tau\Theta,\sigma,t)) = \emptyset =
\mathcal{Z}(\phi,\tau\Theta,\sigma,\alpha(t))$.

Now suppose that $t$ is a constant. Again, we have to distinguish
whether $\sigma = \tau\Theta$ or $\sigma\neq\tau\Theta$. In the first case,
$\alpha(\mathcal{Z}(\phi,\tau\Theta,\sigma,t)) = 
c^\mathcal{A}(\emptyset,\dots,\emptyset) =
\mathcal{Z}(\phi,\tau\Theta,\sigma,\alpha(t))$. In the second case,
$\alpha(\mathcal{Z}(\phi,\tau\Theta,\sigma,t)) = \emptyset =
\mathcal{Z}(\phi,\tau\Theta,\sigma,\alpha(t))$.

Now consider a term $t = \ftau(t_1,\dots,t_n)$ and suppose the result
has been proven for $t_1,\dots,t_n$. Suppose 
$\Nrs(\tau) = \langle\rho_{1},\dots,\rho_{m'}\rangle$ and
$\Rec(\tau) = \langle \rho_{m'+1},\dots,\rho_m \rangle$.
Consider first $\sigma\nrs\phi$. In the following equation sequence,
(*) marks steps that use simple rearrangements such as lifting a
function to sets.
\[
\begin{array}{lr}
\alpha(\mathcal{Z}(\phi,\tau\Theta,\sigma,t)) = & 
\mbox{(Lem.~\ref{subterm-label-lem})}\\
\displaystyle
\alpha\left(
\set{t_i \mid \tau_i\Theta = \sigma} \cup 
\bigcup_{\tau_i\Theta\rec\phi}
\mathcal{Z}(\phi,\tau_i\Theta,\sigma,t_i)\right) = &
\mbox{(*)}\\
\displaystyle
\bigoplus_{\tau_i\Theta = \sigma} \alpha(t_i) \oplus
\bigoplus_{\tau_i\Theta\rec\phi}
\alpha(\mathcal{Z}(\phi,\tau_i\Theta,\sigma,t_i)) = &
\mbox{(ind. hyp.)}\\ 
\displaystyle
\bigoplus_{\tau_i\Theta = \sigma} \alpha(t_i) \oplus
\bigoplus_{\tau_i\Theta\rec\phi}
\mathcal{Z}(\phi,\tau_i\Theta,\sigma,\alpha(t_i)) = &
\mbox{(*)}\\ 
\displaystyle
\bigoplus_{\twocond{\rho_j\Theta = \sigma}{\tau_i = \rho_j}} \alpha(t_i) 
\oplus
\bigoplus_{\twocond{\rho_j\Theta\rec\phi}{\tau_i = \rho_j}}
\mathcal{Z}(\phi,\rho_j\Theta,\sigma,\alpha(t_i)) 
\oplus \\
\multicolumn{1}{r}{\displaystyle
\bigoplus_{\tau_i = \tau} 
\mathcal{Z}(\phi,\tau_i\Theta,\sigma,\alpha(t_i)) =} &
\mbox{(*)}\\
\displaystyle
\bigset{
\bigoplus_{\tau_i = \rho_j} \alpha(t_i) \mid \rho_j\Theta = \sigma}
\cup
\bigcup_{\rho_j\Theta\rec\phi}
\mathcal{Z}\left(\phi,\rho_j\Theta,\sigma,
\bigoplus_{\tau_i = \rho_j} \alpha(t_i)\right)
\oplus\\
\multicolumn{1}{r}{\displaystyle
\mathcal{Z}\left(
\phi,\tau\Theta,\sigma,
\bigoplus_{\tau_i=\tau} \alpha(t_i)\right)
=} &
\mbox{(Lem.~\ref{subterm-label-lem})}
\end{array}
\]

\[
\begin{array}{lr}
\displaystyle
\mathcal{Z}\left(
\phi,\tau\Theta,\sigma,
c^\mathcal{A}
\left(
\bigoplus_{\tau_i=\rho_1} \alpha(t_i),
\dots,
\bigoplus_{\tau_i=\rho_m} \alpha(t_i)
\right)\right)
\oplus\\
\multicolumn{1}{r}{\displaystyle
\mathcal{Z}\left(
\phi,\tau\Theta,\sigma,
\bigoplus_{\tau_i=\tau} \alpha(t_i)
\right)  =} & 
\mbox{(*)}\\
\displaystyle
\mathcal{Z}\left(
\phi,\tau\Theta,\sigma,
c^\mathcal{A}
\left(
\bigoplus_{\tau_i=\rho_1} \alpha(t_i),
\dots,
\bigoplus_{\tau_i=\rho_m} \alpha(t_i)
\right)
\oplus
\bigoplus_{\tau_i=\tau} \alpha(t_i)
\right)  = & 
\mbox{(Def.~\ref{alpha-def})}\\
\mathcal{Z}(\phi,\tau\Theta,\sigma,\alpha(t)).
\end{array}
\]
%PATCH 2
The remaining cases, that either $\sigma\rec\phi$ and
$\sigma\neq\phi$, or $\sigma = \phi$, are very similar and hence omitted.
\end{pf}

We can now state the theorem.

\begin{rmtheorem}\label{alpha-thm}
Let $\tau=c(\vect{u})$ be a flat type such that 
$\Nrs(\tau) = \langle\rho_{1},\dots,\rho_{m'}\rangle$ and
$\Rec(\tau) = \langle \rho_{m'+1},\dots,\rho_m \rangle$.
For any term $t = \ftau(\dots)$, we have
\begin{align*}
\alpha(t)  =_\mathrm{\theory} 
 c^\mathcal{A}\bigg(\! & 
\alpha(\mathcal{Z}(\tau,\tau,\rho_1,t)),
\dots,
\alpha(\mathcal{Z}(\tau,\tau,\rho_{m'},t)),\\[-1ex]
&
\alpha(\mathcal{Z}(\tau,\tau,\rho_{m'+1},t))\cap\mathcal{V},
\dots,
\alpha(\mathcal{Z}(\tau,\tau,\rho_m,t))\cap\mathcal{V}
\bigg)
\oplus 
(\mathcal{Z}(\tau,\tau,\tau,t)\cap\mathcal{V}).
\end{align*}
\end{rmtheorem}

\begin{pf}
Suppose the normal form of $\alpha(t)$ is
$c^\mathcal{A}(a_1,\dots,a_m)\oplus a$.

Consider some $j\in\set{1,\dots,m'}$. Since
 $a$ as well as  $a_k$ for all $m' < k \leq m$, only consist of variables, it
follows by Prop.~\ref{abstract-nrs-rec-prop} and Lemma~\ref{subterm-label-lem} that 
\[
\mathcal{Z}\left(\tau,\tau,\rho_j,c^\mathcal{A}(a_1,\dots,a_m)\oplus a\right)
= a_j.
\]
At the same time, by Lemma~\ref{abst-conc-labelling-lem},
\[
\mathcal{Z}\left(\tau,\tau,\rho_j,\alpha(t)\right) = 
\alpha(\mathcal{Z}\left(\tau,\tau,\rho_j,t)\right).
\]

Now consider some $j\in\set{m'+1,\dots,m}$. Since
 $a$ as well as  $a_k$ for all $m' < k \leq m$, only consist of variables, it
follows by Prop.~\ref{abstract-nrs-rec-prop} and Lemma~\ref{subterm-label-lem} that 
\[
\mathcal{Z}\left(\tau,\tau,\rho_j,c^\mathcal{A}(a_1,\dots,a_m)\oplus a\right)
\cap \mathcal{V}
= a_j.
\]
At the same time, by Lemma~\ref{abst-conc-labelling-lem},
\[
\mathcal{Z}\left(\tau,\tau,\rho_j,\alpha(t)\right) = 
\alpha(\mathcal{Z}\left(\tau,\tau,\rho_j,t)\right).
\]

Finally consider $\tau$. Since
 $a$ as well as  $a_k$ for all $m' < k \leq m$, only consist of variables, it
follows by Prop.~\ref{abstract-nrs-rec-prop} and Lemma~\ref{subterm-label-lem} that 
\[
\mathcal{Z}\left(\tau,\tau,\tau,c^\mathcal{A}(a_1,\dots,a_m)\oplus a\right)
\cap \mathcal{V}
= a.
\]
At the same time, by Lemma~\ref{abst-conc-labelling-lem},
\[
\mathcal{Z}\left(\tau,\tau,\tau,\alpha(t)\right) = 
\alpha(\mathcal{Z}\left(\tau,\tau,\tau,t)\right).
\]
Thus we have shown that the normal form of $\alpha(t)$ is as stated.
\end{pf}

\begin{ex}\label{alpha-Z-ex}
Consider \texttt{LISTS}. We have 
\[
\begin{array}{rlll}
\alpha(\mathtt{[[X],[7]]}) & = &
\mathtt{list^\mathcal{A}
(\alpha(\mathcal{Z}(list(U),list(U),U,[[X],[7]])))}\\
& & \oplus 
\mathtt{\mathcal{Z}(list(U),list(U),list(U),[[X],[7]])\cap\mathcal{V})}\\
& = &
\mathtt{list^\mathcal{A}
(\alpha(\set{[X],[7]}))
\oplus
(\set{[[X],[7]], [[7]], []}\cap\mathcal{V})}\\
& = & 
\mathtt{list^\mathcal{A}
(list^\mathcal{A}(X\oplus int^\mathcal{A}))}
\oplus\emptyset\\
& =_\mathrm{\theory} & 
\mathtt{list^\mathcal{A}
(list^\mathcal{A}(X\oplus int^\mathcal{A}))}.
\end{array}
\]
The theorem tells us how to read the abstract term. First, the absence
of variable on the highest level (i.e.~$\alpha(\mathtt{[[X],[7]]})$ is
not of the form $x\oplus\dots$) means that 
$\mathtt{\mathcal{Z}(list(U),list(U),list(U),[[X],[7]])}$ contains no
variables, or, to use the notation of~\cite{LS01} applied to the
grammar in Ex.~\ref{labelling-ex}, $\zeta(LL,LL,\mathtt{[[X],[7]]})$
is empty. Likewise, the theorem tells us that the argument of the
outermost $\mathtt{list}^\mathcal{A}$ contains the abstraction of all
subterms of $\mathtt{[[X],[7]]}$ returned by 
$\mathtt{\mathcal{Z}(list(U),list(U),U,[[X],[7]])}$, and again in
terms of~\cite{LS01}, the absence of variables at this level tells us
that $\zeta(LL,L,\mathtt{[[X],[7]]})$ is empty.
\end{ex}

\section{The Analysis}\label{analysis-sec}
In this section, we show how an entire program is abstracted based on
an abstraction of the fundamental operation, namely unification. The
abstract program is then given a semantics. This semantics 
{\em describes}, in a well-defined sense, the semantics of the
concrete program. Moreover, under reasonable assumptions, it is
finitely computable. 

\subsection{Abstract Interpretation}
\label{abstract-interpretation-subsec}
In this subsection, we link our formalism to the standard
definitions of abstract interpretation.

Our abstract terms may contain variables, and hence it is only natural
to define substitutions as for concrete terms, only that the range of
those substitutions will contain abstract terms. The instantiation
order $\absleq$ is defined as follows:
$a\absleq b$ if $b\theta^\mathcal{A} =_\mathrm{\theory} a$ for some 
$\theta^\mathcal{A}$. 
It is lifted to substitutions:
$\theta_1^\mathcal{A}\absleq\theta_2^\mathcal{A}$ if 
$a\theta_1^\mathcal{A}\absleq a \theta_2^\mathcal{A}$ for all $a$.
We write $a\approx b$ for $a\absleq b \; \wedge\; b\absleq a$. One should
not confuse $\approx$ with $=_\mathrm{\theory}$! Our notation follows~\cite{CL00},
not~\cite{LS01}. 
An {\bf abstract atom} is an atom using 
abstract terms. We denote the set of abstract atoms
by $\mathcal{B}^\mathcal{A}$.

In order to define and relate semantics for concrete and abstract
programs in the framework of abstract interpretation, we consider sets
of (abstract) atoms with a suitable notion of ordering and
equivalence. We consider the {\em lower power domain} or 
{\em Hoare domain}~\cite{GS90}.
For sets of abstract atoms $I^\mathcal{A}_1$ and
$I^\mathcal{A}_2$, we define 
\[
I^\mathcal{A}_1 \absleq I^\mathcal{A}_2
\; \Leftrightarrow \;
\forall A_1^\mathcal{A} \in I^\mathcal{A}_1 \quad 
\exists A_2^\mathcal{A} \in I^\mathcal{A}_2 \; . \;
A_1^\mathcal{A} \absleq A_2^\mathcal{A}, 
\]
and 
$I^\mathcal{A}_1 \approx I^\mathcal{A}_2$ if 
$I^\mathcal{A}_1 \absleq I^\mathcal{A}_2$ and
$I^\mathcal{A}_2 \absleq I^\mathcal{A}_1$. 
The elements of 
$[2^\mathcal{B^A}]_\approx$ are called {\bf abstract
interpretations}. Abusing notation, we denote 
$[2^\mathcal{B^A}]_\approx$ by $2^\mathcal{B^A}$.

We call a set of abstract atoms $I^\mathcal{A}$ {\bf downwards-closed} 
if $A^\mathcal{A}\in I^\mathcal{A}$ implies 
$\tilde{A}^\mathcal{A}\in I^\mathcal{A}$ for all 
$\tilde{A}^\mathcal{A}\absleq A^\mathcal{A}$. The order relation $\approx$ is 
defined in such a way that each 
$I^\mathcal{A}\in 2^\mathcal{B^A}$ is equivalent to a downwards-closed set. This
observation implies the following lemma.

\begin{rmlemma}\cite[Lemma 3.1]{CL00}
\label{is-lattice-lem}
$(2^\mathcal{B^A},\absleq)$ is a complete lattice.
\end{rmlemma}

\begin{defi}\label{galois-def}
A {\bf Galois insertion} is a quadruple
$\langle (A,\sqsubseteq_A),\alpha,(B,\sqsubseteq_B),\gamma\rangle$
where
\begin{enumerate}
\item
$(A,\sqsubseteq_A)$ and $(B,\sqsubseteq_B)$ are complete lattices of
{\em concrete} and {\em abstract} domains, respectively;
\item
$\alpha: A \rar B$ and $\gamma: B \rar A$ are monotonic functions
called {\em abstraction} and {\em concretisation}, respectively; and
\item
$a \sqsubseteq \gamma(\alpha(a))$ and $\alpha(\gamma(b)) = b$ for
every
$a\in A$ and $b\in B$.
\end{enumerate}
\end{defi}

In the above definition, the $\alpha$ is a priori not the $\alpha$ of
Def.~\ref{alpha-def}, but of course, we have used the same letter
because we link the two in the natural way.

\begin{rmtheorem}\cite[Theorem 3.3]{CL00}
Define $\alpha$ and $\gamma$ as follows:
\[
\begin{array}{ll}
\alpha: 2^\mathcal{B} \rar 2^\mathcal{B^A}, \quad &
\alpha(I) = \set{\alpha(A) \mid A\in I},\\
\gamma: 2^\mathcal{B^A} \rar 2^\mathcal{B}, &
\gamma(I^\mathcal{A}) = \cup \set{I \mid \alpha(I) \absleq
I^\mathcal{A}}.
\end{array}
\]
Then
$\langle 2^\mathcal{B}, \alpha, 2^\mathcal{B^A}, \gamma \rangle$
is a Galois insertion.
\end{rmtheorem}

\begin{defi}\label{describes-def}
An abstract term $a$ {\bf describes} a concrete term $t$, denoted 
$a \describes t$, if $\alpha(t)\absleq a$ (and likewise for atoms).
\end{defi}

Note that for an atom $A$ and an abstract atom $A^\mathcal{A}$, we have
$A^\mathcal{A} \describes A$ if and only if 
$A\in\gamma(\set{A^\mathcal{A}})$. 
For an interpretation $I$ and an abstract interpretation
$I^\mathcal{A}$, we define 
$I^\mathcal{A} \describes I$ if $I\subseteq\gamma(I^\mathcal{A})$, or
equivalently, $\alpha(I) \absleq I^\mathcal{A}$.

\subsection{Abstract Unification}
\label{abstract-unification-subsec}
In this subsection, we show how abstract unification describes
concrete unification. First, we can relate abstraction and application
of a substitution as follows.

\begin{rmlemma}\cite[Lemma 4.1]{CL00}\label{alpha-subst-lem}
Let $t$ be a term an $\theta$ a substitution. Then
$\alpha(t\theta) =_\mathrm{\theory} 
\alpha(t) \set{x/\alpha(x\theta) \mid x\in\dom(\theta)}$.
\end{rmlemma}

\begin{pf}
By structural induction on the term. Let 
$\theta^\mathcal{A} = \set{x/\alpha(x\theta) \mid
x\in\dom(\theta)}$. 

If $t\in\mathcal{V}$ and $t\notin\dom(\theta)$, the result trivially
holds. If $t\in\mathcal{V}$ and $t\in\dom(\theta)$, then
$\alpha(t\theta) = t \set{t/\alpha(t\theta)} = \alpha(t) \theta^\mathcal{A}$.

Now consider $t = \ftau(t_1,\dots,t_n)$, where
$\tau = c(\bar{u})$, 
$\Nrs(\tau) = \langle \rho_1,\dots,\rho_{m'} \rangle$, and
$\Rec(\tau) = \langle \rho_{m'+1},\dots,\rho_m \rangle$, 
 and assume
that the result holds for $t_1,\dots,t_n$. We have
\[
\begin{array}{rcl}
\alpha(t\theta) & = & 
\displaystyle
c^\mathcal{A}
\left(
\bigoplus_{\tau_i=\rho_1} \alpha(t_i\theta),
\dots,
\bigoplus_{\tau_i=\rho_m} \alpha(t_i\theta)
\right)
\oplus
\bigoplus_{\tau_i=\tau} \alpha(t_i\theta) \\
& =  &
\displaystyle
c^\mathcal{A}
\left(
\bigoplus_{\tau_i=\rho_1} \alpha(t_i)\theta^\mathcal{A},
\dots,
\bigoplus_{\tau_i=\rho_m} \alpha(t_i)\theta^\mathcal{A}
\right)
\oplus
\bigoplus_{\tau_i=\tau} \alpha(t_i)\theta^\mathcal{A} = 
\alpha(t)\theta^\mathcal{A}.
\end{array}
\]
\end{pf}

\begin{ex}\label{alpha-subst-ex}
By Lemma~\ref{alpha-subst-lem}, 
\[
\begin{array}{l}
\alpha(\mathtt{[X|Y]}\, \set{\mathtt{X/7, Y/nil}}) =\\ 
\qquad
\alpha(\mathtt{[7]})=\\ 
\qquad
\mathtt{list^\mathcal{A}(int^\mathcal{A})
=_\mathrm{\theory}}\\
\qquad 
\mathtt{list^\mathcal{A}(int^\mathcal{A}) \oplus
list^\mathcal{A}(\emptyset)} =\\
\qquad 
(\mathtt{list^\mathcal{A}(X) \oplus Y})\,
\set{\mathtt{X/int^\mathcal{A}, Y/list^\mathcal{A}(\emptyset)}} =
\qquad \\ 
\multicolumn{1}{r}{
\alpha(\mathtt{[X|Y]})\,
\set{\mathtt{X/\alpha(7), Y/\alpha(nil)}}}.
\end{array} 
\]
\end{ex}

The following theorem is a straightforward consequence.

\begin{rmtheorem}\cite[Thm.~4.2]{CL00}\label{alpha-order-thm}
Let $t_1, t_2$ be terms. If $t_1\leq t_2$ then 
$\alpha(t_1) \absleq \alpha(t_2)$ (and likewise for atoms).
\end{rmtheorem}

\begin{defi}\label{unifier-def}
We denote by 
$cU_\mathrm{\theory}(o_1,o_2)$ a complete set of 
\theory-unifiers of syntactic objects $o_1, o_2$, i.e., a set of
abstract substitutions such that for each 
$\theta^\mathcal{A}\in cU_\mathrm{\theory}(o_1,o_2)$, we have
$o_1\theta^\mathcal{A} =_\mathrm{\theory} o_2\theta^\mathcal{A}$, and for 
any $\tilde{\theta}^\mathcal{A}$ such that 
$o_1\tilde{\theta}^\mathcal{A}=_\mathrm{\theory} o_2\tilde{\theta}^\mathcal{A}$, 
we have $\tilde{\theta}^\mathcal{A}\absleq \theta^\mathcal{A}$ for some 
$\theta^\mathcal{A}\in cU_\mathrm{\theory}(o_1,o_2)$.
\end{defi}

The next theorem says that unification of abstract terms is a correct
abstract unification. The second part of the statement says that an abstract
substitution corresponding to a concrete substitution as stated
correctly mimics that concrete substitution, for any atom.

\begin{rmtheorem}\cite[Thm.~4.4]{CL00}\label{abstract-unification-correct-thm}
Let $A_1, A_2$ be atoms that are unifiable with MGU $\theta$,
and $A_1^\mathcal{A},A_2^\mathcal{A}$ be abstract atoms such that   
$A_1^\mathcal{A}\describes A_1$ and $A_2^\mathcal{A}\describes A_2$.
Then there exists a unifier 
$\theta^\mathcal{A}\in
cU_\mathrm{\theory}(A_1^\mathcal{A},A_2^\mathcal{A})$ such that 
$A_1^\mathcal{A}\theta^\mathcal{A} \describes A_1\theta$, and moreover
for any atom $B$, we have 
$\alpha(B)\theta^\mathcal{A} \describes B\theta$.
\end{rmtheorem}

\begin{pf}
Consider the pairs $\langle B,A_1\rangle$, $\langle B,A_2\rangle$
where $B$ is an arbitrary atom. Since 
$\langle B,A_i\rangle\theta\leq\langle B,A_i\rangle$, we have by
Thm.~\ref{alpha-order-thm} and the definition of $\describes$
\[
\alpha(\langle B, A_i \rangle\theta)\absleq
\alpha(\langle B, A_i \rangle)\absleq
\langle \alpha(B),A_i^\mathcal{A}\rangle
\quad (i = 1,2).
\]
Since $A_1\theta=A_2\theta$, it follows that
$\alpha(\langle B, A_1 \rangle\theta) = 
 \alpha(\langle B, A_2 \rangle\theta)$ and so 
$\alpha(\langle B, A_1 \rangle\theta)$ is a common
\theory-instance of   
$\langle \alpha(B),A_1^\mathcal{A}\rangle$ and
$\langle \alpha(B),A_2^\mathcal{A}\rangle$. Hence
\[
cU_\mathrm{\theory}(\langle \alpha(B), A_1^\mathcal{A}\rangle,
                    \langle \alpha(B), A_2^\mathcal{A}\rangle)
\]
contains a $\theta^\mathcal{A}$ such that 
$\alpha(\langle B, A_1 \rangle\theta) \absleq
\langle \alpha(B), A_1^\mathcal{A}\rangle\theta^\mathcal{A}$ and so
$\alpha(B\theta) \absleq \alpha(B)\theta^\mathcal{A}$ and
$\alpha(A_1 \theta) \absleq A_1^\mathcal{A}\theta^\mathcal{A}$. 
Now
$cU_\mathrm{\theory}(\langle \alpha(B), A_1^\mathcal{A}\rangle,
                    \langle \alpha(B), A_2^\mathcal{A}\rangle)$
is also a complete set of unifiers of $A_1^\mathcal{A}$ and
$A_2^\mathcal{A}$, and so the claim follows.
\end{pf}

In addition, we also have that \theory-unification is optimal. To make
this notion precise, consider two abstract atoms $A_1^\mathcal{A}$,
$A_2^\mathcal{A}$, and let 
$I_i = \gamma(\set{A_i^\mathcal{A}})$ ($i = 1,2$). For 
two atoms $A_1\in I_1$, $A_2\in I_2$, any common instance is in
$I_1\cap I_2$, since $I_1$, $I_2$ are downwards-closed. 
Now let 
$cI_\mathrm{\theory}(A_1^\mathcal{A},A_2^\mathcal{A}) = 
\set{A_1^\mathcal{A} \theta^\mathcal{A} \mid 
\theta^\mathcal{A} \in cU_\mathrm{\theory}(A_1^\mathcal{A},A_2^\mathcal{A})}$. 
We might call $cI_\mathrm{\theory}(A_1^\mathcal{A},A_2^\mathcal{A})$ a
{\em complete} set of common \theory-instances of 
$A_1^\mathcal{A}$, $A_2^\mathcal{A}$. Optimality of abstract
unification means that  
$cI_\mathrm{\theory}(A_1^\mathcal{A},A_2^\mathcal{A})$ describes only
the atoms in $I_1\cap I_2$.

\begin{rmtheorem}\cite[Thm.~4.6]{CL00}
\label{abstract-unification-optimal-thm}
For $i = 1,2$, let $A_i^\mathcal{A}$, be abstract atoms and 
$I_i = \gamma(\set{A_i^\mathcal{A}})$.
Then $cI_\mathrm{\theory}(A_1^\mathcal{A},A_2^\mathcal{A})$ describes only
the atoms in $I_1\cap I_2$.
\end{rmtheorem}

\begin{pf}
To derive a contradiction, assume that $B$ is an atom such that for
some
$B^\mathcal{A}\in
cI_\mathrm{\theory}(A_1^\mathcal{A},A_2^\mathcal{A})$, 
we have $B^\mathcal{A}\describes B$ but $B\not\in I_1 \cap I_2$. This
implies that $A_1^\mathcal{A}\not\describes B$ or
$A_2^\mathcal{A}\not\describes B$. On the other hand, $B^\mathcal{A}$
is a common instance of $A_1^\mathcal{A}$, $A_2^\mathcal{A}$, which
implies $A_1^\mathcal{A}\describes B$ and
$A_2^\mathcal{A}\describes B$. Contradiction.
\end{pf}

\subsection{Abstraction of Programs}\label{program-abstraction-subsec}
In~\cite{LS01,SHK00}, programs were assumed to be in normal, also
called canonical, form. In~\cite{LS01}, the abstraction of a
unification constraint $x = y$ involves computing the intersection of
the two grammars corresponding to $x$ and $y$. In addition, the
abstraction of a unification constraint $y = f(x_1,\dots,x_n)$ involves
computing a grammar for $f(x_1,\dots,x_n)$ from the $n$ grammars for
$x_1,\dots,x_n$. In~\cite{SHK00}, no such operations are performed,
but still the abstraction of unification constraints is not obvious. 
Each unification constraint $y = f(x_1,\dots,x_n)$ is abstracted
as a call $\fdep(y,x_1,\dots,x_n)$, where $\fdep$ is a predicate that
expresses the relationship between the abstraction of a term and its
subterms. 

In the framework we have set up here following~\cite{CL00},
abstracting a program is much simpler. We have designed the domains
so that Thm.~\ref{abstract-unification-correct-thm} holds, and so we can
abstract a program simply by replacing each term with its abstraction.
Thus $\alpha$ is lifted in the obvious way to atoms, clauses, programs
and queries.

The semantics of the abstract program will be defined using an 
\theory-enhanced version of the $T_P$-operator. Formally
\begin{align*}
T^\mathcal{A}_P(I^\mathcal{A}) = 
\set{\alpha(H)\theta^\mathcal{A} \mid & 
C = H \lar B_1,\dots,B_n \in P,
\langle A^\mathcal{A}_1,\dots,A^\mathcal{A}_n \rangle \renamed{C} I^\mathcal{A},\\
&
\theta^\mathcal{A}\in 
cU_\mathrm{\theory} 
(\langle \alpha(B_1),\dots,\alpha(B_n)\rangle,\langle A^\mathcal{A}_1,\dots,A^\mathcal{A}_n\rangle)}.
\end{align*}

We denote by $\sem{P^\mathcal{A}}_\mathrm{\theory}$ the 
least fixpoint of $T^\mathcal{A}_P$, which exists
by~\cite[Cor.~5.2]{CL00}.  
The following theorem says that this abstract semantics correctly
describes the concrete semantics. 

\begin{rmtheorem}\cite[Thm.~5.4]{CL00}\label{correct-semantics-thm}
Let $P$ be a program. Then 
$\sem{\alpha(P)}_\mathrm{\theory}
\describes
\sem{P}_s
$.
\end{rmtheorem}

% PATCH 3

We could make further statements about the semantics, e.g.~call and
answer patterns; for that purpose, we would use the magic set
transformation. However, those results would be completely along the
lines of~\cite{CL96,CL00}, as are the proofs of the results we gave in
this subsection.

\subsection{Finiteness}\label{finiteness-subsec}
In~\cite{CL00} we find a result that the abstract semantics of a
program is finite provided that the type abstraction is
monomorphic. 
The result does not hold anymore for polymorphic type
abstractions, and the authors give the program 
\[
P_1 :=
\set{
\mathtt{p([X]) \lar p(X).},\ 
\mathtt{p(1).}}
\] 
as an example. As a solution, the authors propose a 
{\em depth-$k$ abstraction}, i.e., some ad-hoc bound on the depth of
types. 

It is understandable that a descriptive view of types leads to the
conviction that infinity of the abstract semantics is inherent in a
polymorphic type abstraction and cannot reasonably be
avoided. 

Instead of $P_1$, consider the program
\[
P_2 := \set{
\mathtt{p([X]) \lar p(X).},\ 
\mathtt{p([]).}}
\] 
For the argument in~\cite{CL00}, $P_2$ is completely equivalent
to $P_1$, i.e., the authors could just as well have chosen
$P_2$. However, using $P_2$ allows us to make a stronger point than using
$P_1$. The reason is that both programs are not typable
by the rules of Table~\ref{rules-tab} (that is to say, in a
prescriptive approach to typing), but for $P_2$, this is much less
obvious.
The program $P_2$ is forbidden due to the {\em head
condition}~\cite{HT92}, i.e., the special typing rule $\mathit{Head}$
which is different from rule $\mathit{Atom}$ (see
Table~\ref{rules-tab}). 

\begin{propo}\label{head-propo}
Assuming $\mathcal{F}_\mathtt{LISTS}$ (see
Ex.~\ref{typerelations-ex}), there exists no variable typing
$\Gamma$ such that 
$\Gamma \vdash \mathtt{p([X]) \lar p(X)}\ \mathit{Clause}$,
regardless of what the declared type of $\mathtt{p}$ is.
\end{propo}

\begin{pf}
To derive
$\Gamma \vdash \mathtt{p([X]) \lar p(X)}\ \mathit{Clause}$, we need to 
derive
$\Gamma \vdash \mathtt{p([X])}\ \mathit{Head}$, and in turn,
\begin{equation}
\label{list-x-has-type-eq}
\Gamma \vdash \mathtt{[X]}:\sigma
\end {equation}
for some type $\sigma$. 
By rule $\mathit{Func}$ and the fact that the declared range type of
$\mathtt{Cons}$ is $\mathtt{List(U)}$, it follows that 
$\sigma = \mathtt{List(\tau)}$ for some $\tau$, and so for 
$\Gamma \vdash \mathtt{p([X])}\ \mathit{Head}$ to be a valid type
judgement, the declared type of $\mathtt{p}$ must be 
$\mathtt{List(\tau)}$, so we write $\mathtt{p}_\mathtt{List(\tau)}$. 

To derive
$\Gamma \vdash \mathtt{p([X]) \lar p(X)}\ \mathit{Clause}$, we 
{\em also} need to derive
$\Gamma\vdash \mathtt{p_{list(\tau)}(X)}: \mathit{Atom}$, and in turn,
$\Gamma\vdash \mathtt{X}:\mathtt{list}(\tau)\Theta$ for some type
substitution $\Theta$. This implies that 
$\mathtt{X}:\mathtt{list}(\tau)\Theta \in \Gamma$ and hence
$\Gamma\vdash \mathtt{[X]}:\mathtt{list(list(\tau))}\Theta$, and in
particular,
$\Gamma\not\vdash \mathtt{[X]}:\mathtt{list(\tau)}$. This is a
contradiction to (\ref{list-x-has-type-eq}), showing that there exists 
no $\Gamma$ such that 
$\Gamma \vdash \mathtt{p([X]) \lar p(X)}\ \mathit{Clause}$.
\end{pf}

We want to show that disregarding such programs, the abstract
semantics is always finite. We first need the following lemma about
concrete programs, stating that the arguments of any
atom in $\sem{P}_s$ are of the declared type.

\begin{rmlemma}\label{finitely-many-types-lem}
Let $P$ be a typed program. For any atom
$\ptau(t_1,\dots,t_n)\in\sem{P}_s$, we have
$\_\vdash (t_1,\dots,t_n):(\tau_1,\dots,\tau_n)$.
\end{rmlemma}

\begin{pf}
Suppose $I$ is a set if atoms having the property stated for
$\sem{P}_s$. We show that an application of the $T_P$-operator to 
$I$ preserves the property. This immediately implies the result.

Consider some clause
$C = \ptau(t_1,\dots,t_n) \lar B_1,\dots,B_m$ in $P$, and suppose that
$\langle A_1,\dots,A_m \rangle \renamed{C} I$, such that 
$\theta = 
MGU(\langle B_1,\dots,B_m\rangle,\langle A_1,\dots,A_m\rangle)$.
 By the rules in
Table~\ref{rules-tab}, in particular $\mathit{Head}$, we have
$\_\vdash (t_1,\dots,t_n):(\tau_1,\dots,\tau_n)$. 
For each $B_l = q_{\sigma_1,\dots,\sigma_{n'}}(s_1,\dots,s_{n'})$, 
by the rules in Table~\ref{rules-tab}, we have
$\_\vdash (s_1,\dots,s_{n'}): (\sigma_1,\dots,\sigma_{n'})\Theta$ for
some $\Theta$. Let $A_l = q(r_1,\dots,r_{n'})$. 
By assumption about $I$, 
we have 
$\_\vdash (r_1,\dots,r_{n'}): (\sigma_1,\dots,\sigma_{n'})$ and hence, 
by the typing rules, also 
$\_\vdash (r_1,\dots,r_{n'}): (\sigma_1,\dots,\sigma_{n'})\Theta$.
By standard results~\cite[Thm.~1.4.1, Lemma 1.4.2]{HT92}, it follows that  
$\_\vdash (s_1,\dots,s_{n'})\theta:
(\sigma_1,\dots,\sigma_{n'})\Theta$.
Since the choice of $A_l$ was arbitrary, it follows that each atom in
$C\theta$ can be typed using the same types as for the corresponding
atom in $C$. This applies in particular for the clause head, and so
$\_\vdash (t_1,\dots,t_n)\theta:(\tau_1,\dots,\tau_n)$.
\end{pf}

By Prop.~\ref{abstract-term-is-typed-prop}, the above lemma
applies also to the abstraction of a program.

\begin{coro}\label{finitely-many-types-cor}
Let $P$ be a typed program. For any atom
$\ptau(a_1,\dots,a_n)\in\sem{\alpha(P)}_\mathrm{\theory}$, we have
$\_\vdash^\mathcal{A} (a_1,\dots,a_n):(\tau_1,\dots,\tau_n)$.
\end{coro}

The following lemma states that for a given type $\phi$, 
there are only finitely many different abstract terms for $\phi$.

\begin{rmlemma}\label{finitely-many-terms-lem}
For any type $\phi$, the set of abstract terms
$\set{ a \mid \_\vdash^\mathcal{A} a:\phi}$ is finite modulo $\approx$.
\end{rmlemma}

\begin{pf}
Since the claim is that the set is finite modulo $\approx$, it is
clearly sufficient to restrict our attention to {\em normal} abstract
terms. It is useful to recall the notations and definitions of
Subsec.~\ref{normal-abstract-terms-subsec}. 

The proof is along the lines of the proof of~\cite[Thm.~3.2]{CL00},
but matters are slightly more complicated since our abstract terms are 
nested. 

We define by structural induction:
\begin{itemize}
\item
for each abstract term $a\oplus x^\oplus$, $\epsilon$ is a {\em path},
and any variable in $x^\oplus$ is a variable 
{\em occurring in $a\oplus x^\oplus$ at $\epsilon$}; 
\item
for an abstract term $c^\mathcal{A}(a_1,\dots,a_m)\oplus\dots$, if 
$\zeta$ is a path for $a_j$ and $x$ is a variable occurring in $a_j$
at $\zeta$, then $j.\zeta$ is a 
{\em path for $c^\mathcal{A}(a_1,\dots,a_m)\oplus\dots$}, 
and $x$ is a variable
{\em occurring in $c^\mathcal{A}(a_1,\dots,a_m)\oplus\dots$ at $j.\zeta$}.
\end{itemize}
By the well-definedness of normal abstract terms, it follows that that 
there is a maximum number of paths that a normal abstract term for
$\phi$ can have. Let $n$ be this number for our given $\phi$.

Suppose that a normal abstract term $a$ for $\phi$ 
contains more that $2^n-1$ distinct variables. By a simple
combinatorial argument, one sees that there must be at least two
variables, say $x$ and $y$, occurring at exactly the same paths in
$a$. Consider $a' = a \set{x/z, y/z}$. Trivially
$a' \absleq a$. On the other hand, we have
$a' \set{z/x\oplus y} =_\mathrm{\theory} a$, and thus
$a \absleq a'$. So we have $a \approx a'$, and 
$\#(\vars(a')) = \#(\vars(a)) - 1$. 

By iterating this argument, it follows that any normal abstract term
for $\phi$ is $\approx$-equivalent to a term containing no more that
$2^n-1$ variables, and thus the claim follows.
 \end{pf}

The following is a simple corollary.

\begin{coro}\label{finitely-many-atoms-cor}
Let $\ptau$ be a predicate and $\Theta$ a type substitution.
Modulo $\approx$, there are only finitely many abstract atoms
$p(a_1,\dots,a_n)$ such that $(a_1,\dots,a_n)$ is a vector of 
normal abstract terms for the type vector 
$(\tau_1,\dots,\tau_n)\Theta$.
\end{coro}

\begin{rmtheorem}\label{finite-semantics-thm}
Let $P$ be a typed program. Then 
$\sem{\alpha(P)}_\mathrm{\theory}$ is finite.
\end{rmtheorem}

\begin{pf}
By Corollaries~\ref{finitely-many-types-cor}
and~\ref{finitely-many-atoms-cor}. 
\end{pf}

As it stands, the theorem depends critically on the fact that we
assume a {\em bottom-up} semantics. To explain this, consider the
program 
\[
P_3 = 
\set{
\mathtt{p(X) \lar p([X]).},\ 
\mathtt{p([]).}},
\] 
which at first look is very similar to the program 
$P_2$
given at the beginning of this subsection. However, assuming that
$\mathtt{p}$ has declared type $\mathtt{list(U)}$, the program $P_3$ is 
typed according to the rules of Table~\ref{rules-tab}. Therefore, of
course, Thm.~\ref{finite-semantics-thm} applies to this program. 

Note that $P_3$, when called with the query $\mathtt{p(Y)}$, gives rise 
to infinitely many calls 
$\mathtt{p(Y)}, 
\mathtt{p([Y])}, 
\mathtt{p([[Y]])},\dots$, with abstractions
$\mathtt{p(Y)}, 
\mathtt{p(list^\mathcal{A}(Y))},\allowbreak 
\mathtt{p(list^\mathcal{A}(list^\mathcal{A}(Y)))},\allowbreak
\dots$.
So the set of calls cannot be described (in the technical sense,
using $\describes$) finitely.  

We make two observations about $P_3$:
\begin{itemize}
\item
The {\em magic set} version~\cite{CD95} of the program contains the
clause 
\[
\mathtt{p^c([X]) \lar p^c(X).},
\] which is to be read as ``in
order for $\mathtt{p([X])}$ to be called, $\mathtt{p(X)}$ must be
called. This clause is not typable according to the rules of
Table~\ref{rules-tab} as the head condition is violated. Thus
Thm.~\ref{finite-semantics-thm} is not applicable. This is not
surprising given the fact that the very purpose of the magic set
transformation is to characterise {\em calls}.
\item
In the literature on prescriptive typing, the
behaviour exhibited by $P_3$ has been called 
{\em polymorphic recursion}~\cite{H93}.
It is by no means common. In fact, it 
is very much on the borderline of what is allowed in prescriptively
typed programming languages. In ML for example, it is forbidden, as it 
breaks the capabilities of the type inference procedure.
\end{itemize}
It has previously been suspected that there is an interesting
relationship between the head condition and polymorphic recursion, which 
deserves some profound investigation~\cite{DS01}. The two observations 
above add weight to this.

In this paper, we do not want to study the difference between top-down
and bottom-up semantics in detail. Nevertheless, we now formulate what
it means for a program not to use polymorphic recursion, in which case
we say that it only uses monomorphic recursion. 

For simplicity, we assume that a program 
only contains direct recursion, i.e., no recursion of the form
$p \lar \dots q$,  $q \lar \dots p$. It is straightforward to
generalise our considerations otherwise. In the following, we use
$\vect{t}$ ($\vect{\tau}$) to denote vectors of terms (types).

\begin{defi}\label{monomorphic-recursion-def}
A typed program {\bf uses only monomorphic recursion} if for any clause
$p(\vect{t}) \lar \dots p(\vect{s}) \dots$, we have
$\_\vdash \vect{t}:\vect{\tau},\ \vect{s}:\vect{\tau}$
for some vector of types $\vect{\tau}$.
\end{defi}

Thus monomorphic recursion means that in each clause, the type of any
recursive call must be identical to the type of the head.
Since the head condition is still in force, this implies that any
recursive call must have a type which is identical to the declared
type of its predicate.

For programs using only monomorphic recursion, it should be possible
to devise a variant of the magic set transformation such that the
transformed program is typed according to the rules in
Table~\ref{rules-tab}, and therefore has a finite abstract semantics.

\section{Towards an Implementation}
\label{implementation-sec}
So far, we have not implemented the analysis proposed in this
paper. As far as computing the semantics of the abstract program is
concerned, the only difference with the implementations mentioned
in~\cite{CL96,CL00} is that instead of ACI or ACI1 we have the
equality theory \theory. The former theories are finitary~\cite{BS98},
and the corresponding unification 
problems are NP-complete. Obviously, \theory\ cannot behave any
better. Studying \theory\ is a topic for future work, but we would
certainly expect it to be finitary as well. 

There is an implementation of the analysis we proposed
in~\cite{SHK00}, which essentially aims at the same degree of
precision we have here, but the framework is different. In fact, this
paper relates to~\cite{SHK00} in the same way as~\cite{CL96,CL00} relates
to~\cite{CD94}. This is interesting because the authors mention that
an implementation using ACI-unification turned out to be much 
faster than the implementation in~\cite{CD94}.

Note that to compute the abstraction of a program,
in~\cite{CD94,CL00}, it is the user who has to provide information
about the particular type language used in a program (see paragraph
after Def.~\ref{alpha-def}), whereas in our analysis, this information
is extracted from the declared types. We had previously shown~\cite{SHK00} that
analysing the type declarations (computing the NRSs and recursive
types) is viable even for some contrived, complex type declarations,
which one would never expect in practice, since {\em good programs
have small types}~\cite{H93}\footnote{However, this should not be
understood as a contradiction to our \texttt{NESTS} or \texttt{TABLES}
examples. Henglein comes from a functional programming background, 
and in that community, those type declarations would by all means
still qualify as ``small''.}.
 
In~\cite{CL96}, there is some speculation as to why abstract
unification is not as bad as it seems by the theoretical result that
it is NP-complete. It is said that usually unifications involve
``few'' variables. Here we want to substantiate that claim somewhat,
but it remains speculation all the same. 
\begin{itemize}
\item
The first argument is that abstract terms (in normal form) are likely
to be linear. Recall that the abstract terms are designed in such a
way that different positions correspond to different subterm
types. Since we use prescriptive types, one is tempted to conclude
that abstract terms {\em must} be linear, since the same variable
cannot have different types. That however would be a fallacy, since
via instantiation of types, different subterm types can become
equal. For example, the term
$\mathtt{node(null,X,X,eq,null)}$ of type 
$\mathtt{table(string)}$ has the abstraction
$\mathtt{table^\mathcal{A}(X,bal^\mathcal{A},X)}$ which is not linear.
Nevertheless, this should be an exception.
\item
On the whole, logic programs are based on a very simple notion of
modes. Of course, it is the very exceptions to this rule that justify
developing a complex instantiation analysis like the one of this paper
or~\cite{CL00,LS01}, but still, often one deals with simple
assignments rather than full unification in concrete programs, and
this carries through to abstract programs. 
\end{itemize}

To give at least one example of the advance of our analysis
over~\cite{CL00}, we use $\mathtt{table(int)}$. Suppose there is a
predicate $\mathtt{insert}/4$ whose arguments represent:  a table $t$, 
a key $k$, a value $v$, and a table
obtained from $t$ by inserting the node whose key is $k$ 
and whose value is $v$.
From the abstract semantics of the program, it is possible 
to read that a query whose abstraction is 
\[
\mathtt{insert(
table^\mathcal{A}(int^\mathcal{A},bal^\mathcal{A},str^\mathcal{A}),
str^\mathcal{A},V2,T)},
\] 
i.e., a query to insert 
an {\em uninstantiated} value into a ground table, yields an answer
whose abstraction is 
\[
\mathtt{insert(
table^\mathcal{A}(int^\mathcal{A},bal^\mathcal{A},str^\mathcal{A}),
str^\mathcal{A},V2,
table^\mathcal{A}(int^\mathcal{A}\oplus V2,bal^\mathcal{A},str^\mathcal{A}))},
\] 
i.e., the result is a table whose values may be uninstantiated. 

%\section{Groundness Analysis}\label{groundness-sec}

%\section{Sharing Analysis}\label{sharing-sec}

\section{Discussion}\label{discussion-sec}
In this paper, we have proposed a formalism for deriving abstract
domains from the type declarations of a typed program. Effectively, we
have recast our previous work~\cite{SHK00} using
important parts of the formalisms of~\cite{CL00,LS01}.
We now compare this paper with those two works under several aspects.

\paragraph{The type system.}
Using the terminology introduced in this paper, one can say
that~\cite{CL00} uses a polymorphic type system with the following
assumptions: types are either monomorphic or unary, and
the only subterm types of a unary type $c(u)$ are $c(u)$ itself 
(and $c(u) \rec c(u)$) and $u$ (and $u \nrs c(u)$). This is the 
simplest thinkable scenario of proper polymorphism; in fact, 
only lists and trees are covered. Our \texttt{TABLES} or let alone the 
\texttt{NESTS} example are not covered. In contrast,~\cite{LS01}
assumes regular types without polymorphism. Thus there are only
finitely many types the analysis has to deal with. However, those may
be very complex; e.g., one can easily construct a grammar that
corresponds to the type $\mathtt{table(int)}$. So the type systems
of~\cite{CL00,LS01} are not formally comparable, but the type system
we assume in this paper is a strict generalisation of both.

\paragraph{Descriptive vs.~prescriptive types.}
According to the authors' claims,~\cite{CL00} takes a
descriptive view of typing, whereas~\cite{LS01} takes a prescriptive
view. However, we find that the formalism of~\cite{CL00} can very well
be adapted to prescriptive typing. On the other hand, we find that
some aspects of~\cite{LS01} belong rather to a descriptive view of
typing. 

First, the fact that the typing approach of~\cite{CL00} is descriptive
rightly accounts for the fact that they must consider ``ill-typed''
terms such as $\mathtt[1|2]$. In this paper, all terms are
``well-typed'', and so are the abstract terms. 

In~\cite{LS01} it is assumed that a unique type (or equivalently, grammar) is associated
with each program variable. A unification constraint in a program
gives rise to operations such as computing the intersection of two
types and computing the type of a term from the types of its
subterms. Such operations can improve the precision of an analysis,
e.g.~if $\mathtt{X}$ has the declared type 
``list with an even number of elements'' and 
$\mathtt{Y}$ has the declared type 
``list with a number of elements divisible by $3$'', then a unification 
$\mathtt{X} = \mathtt{Y}$ implies that both variables have the type 
``list with a number of elements divisible by $6$''. 
In our opinion, the presence of such operations introduces an aspect
of type {\em inference} into their formalism which is somewhat in
contradiction  to a {\em prescriptive} 
approach to typing. While such type inference may be useful, the
authors do not give a convincing example for it being so.

\paragraph{Labellings.}
Labellings are useful to formalise which aspects of the structure of a
concrete term we want to capture by our analysis, and so it is natural
that we used them mainly in Sec.~\ref{structure-sec}. In~\cite{CL00},
they are absent, although they may have been useful (see
Sec.~\ref{alpha-label-relation-sec}). In~\cite{SHK00}, there
were similar functions called {\em extractors} and {\em termination functions}.

First note that $\zeta$ only collects variables, whereas $\mathcal{Z}$ 
also collects non-variable terms. This generalisation allows us to
describe the relation between a term and its abstraction as we did in
Sec.~\ref{alpha-label-relation-sec}. 

The labelling function $\zeta$ in~\cite{LS01} has three arguments: a
grammar (which however can be identified with its starting
non-terminal), a non-terminal to be labelled, and a labelling
term. Our labelling function $\mathcal{Z}$ has {\em four} arguments. 
We found it useful to have as first argument a {\em flat} type
(e.g.~$\mathtt{nest(V)}$) which
gives us a certain grammar, but also allow for productions of that
grammar starting from some other non-terminal 
(e.g.~$\mathtt{list(nest(V))}$). Actually, it may well be the 
case that the first argument is redundant, i.e.~that the grammar can 
be derived from the starting non-terminal (e.g.~$\mathtt{nest(V)}$
from $\mathtt{list(nest(V))}$). We prefer however to keep this useful
intuitive explanation: one argument to indicate the grammar, one to 
indicate the starting non-terminal.
The difference between our labelling function and that of~\cite{LS01}
is due to polymorphism.

\paragraph{Abstract terms.}
 In~\cite{LS01}, the abstraction of terms
is not actually made explicit, but effectively, given a program
variable $x$, its abstraction is the (somehow ordered) tuple of 
non-terminals of the grammar of $x$. Non-terminals are thought of as
{\em abstract variables} (see the paragraph after Equation
(\ref{aci1-eq})). 
Our abstraction of terms, denoted $\alpha$, is designed in such a way
that the abstraction $\mathit{type}$ in~\cite{CL00} is essentially a
special case of it. We do not introduce abstract variables but rather 
collect the labellings for a term in a structure called 
{\em abstract term}. The reason for this decision is that it
allows us to deal with arbitrarily large grammars/type graphs.

\paragraph{Type hierarchies.}
Given a function $f_{\dots\rar c(u)}$, 
the abstraction $\mathit{type}$ in~\cite{CL00} distinguishes between 
the argument positions of declared type $u$ and the ``recursive''
argument positions. Via type instantiation, this already gives rise to
a certain hierarchy of arbitrary depth, as reflected for example in 
the abstract term 
$\mathtt{list^\mathcal{A}(list^\mathcal{A}(int^\mathcal{A}))}$.
Our concepts of {\em non-recursive subterm type} and 
{\em recursive type} generalise this idea. An NRS of a flat type is
not necessarily a parameter, and $\tau$ can have other recursive types
than $\tau$ itself. This hierarchy allows us to deal with the fact 
that through instantiation of types, a polymorphic language gives rise 
to arbitrarily big type graphs (grammars).

In contrast, in~\cite{LS01}, all non-terminals (types) reachable from
the starting node of a grammar are treated in the same way. This
approach is viable since the size of the grammars is fixed beforehand.

\paragraph{Equality theory.}
The equality theory for evaluating abstract terms in~\cite{LS01} is
ACI1. Distributivity is not applicable.  In~\cite{CL00}, the equality
theory is ACI, so there is no neutral element. The authors mention
distributivity but decide against it. This is in contrast
to~\cite{CL96} where the equality theory is ACI1 plus distributivity.
Our extraction axioms are only
relevant for a language where some type has a recursive type other
than itself, and so it is not applicable to~\cite{CL96,CL00}.
We believe that at least conceptually, both a neutral element and the
distributivity axioms are very natural, even if at the level of an
implementation, it might be preferable not to have them.

\paragraph{Types = abstract terms?}
In~\cite{CL00}, there is no distinction between a type constructor $c$
and the function $c^\mathcal{A}$ to build abstract terms. Also, the
equivalent of our function $\alpha$ is
called {\em type abstraction} and denoted by $\mathit{type}$, which
highlights the fact that in descriptive approaches to typing, 
{\em type} analysis and {\em mode} analysis are blurred almost to the
extent of being considered to be the same thing. However, such an
identification only works because the assumptions about the type
system are so restrictive. \\

\noindent
Thus we have generalised~\cite{CL00,LS01} by considering a type system
which almost (see below) corresponds to the type system of existing typed
programming languages.  We have given several examples in
Sec.~\ref{structure-sec} hoping to convince the reader that such a
generalisation is non-trivial. In particular, there are two natural
requirements: the construction of an abstract domain for a polymorphic
type should be truly parametric, and the abstract domains should be
finite for a given program and query. We had to impose two conditions
on the type declarations to ensure these requirements.
On a technical level, the fact that the SCCs of a type graph are not
stable under instantiation makes it difficult to meet those
requirements.

We now very briefly recall some other related work. We refer to the
discussions in~\cite{CL00,LS01,SHK00} for more details.

Both this paper and~\cite{CL00,LS01} build on ideas presented
originally in~\cite{CD94}. 

{\em Recursive modes}~\cite{TL97} characterise that
the left spine, right spine, or both, of a term are instantiated.
This seems ad-hoc but often coincides with characterising that
all recursive subterms of a term are instantiated.

A system for type analysis of Prolog is presented in~\cite{HCC93}. It
takes a descriptive approach to typing, and the abstract domains are,
in general, infinite. Therefore, widenings must be used. 
Similarly, in~\cite{JB92},  the finiteness of
abstract domains and terms is ensured by imposing an ad-hoc
bound on the number of symbols.

It would probably be possible to express the abstraction of terms
proposed here as application of a particular 
{\em pre-interpretation}~\cite{GBS95}. 

A classical instantiation analysis is not interesting for
Mercury~\cite{mercury} as the language is strongly moded. However,
our work might also have applications for Mercury.

\subsection*{Acknowledgements}
The author was supported by the ERCIM fellowship programme.

\bibliography{modes_types,thesis}

\appendix
%\section{Proofs}\label{proofs-sec}
%We first show the following simple statement:
%$\alpha(t) = \alpha(\mathcal{Z}(\tau,\tau,\tau,t))$. The proof is by
%structural induction. The base case 
%$\mathcal{Z}(\tau,\tau,\tau,t) = t$ is trivial. In the inductive case,
%we have
%\[
%\begin{array}{rcl}
%\alpha(t) & = &  
% c^\mathcal{A}
%\left(
%\bigoplus_{\tau_i=\rho_1} \alpha(t_i),
%\dots,
%\bigoplus_{\tau_i=\rho_m} \alpha(t_i)
%\right)
%\oplus
%\bigoplus_{\tau_i=\tau} \alpha(t_i) \\
%& = &  
% c^\mathcal{A}
%\left(
%\bigoplus_{\tau_i=\rho_1} \alpha(t_i),
%\dots,
%\bigoplus_{\tau_i=\rho_m} \alpha(t_i)
%\right)
%\oplus
%\bigoplus_{\tau_i=\tau} \alpha(t_i)
%\oplus
%\bigoplus_{\tau_i=\tau} \alpha(t_i) \\
%& = &  
%\alpha(t)
%\oplus
%\bigoplus_{\tau_i=\tau} \alpha(\mathcal{Z}(\tau,\tau,\tau,t_i)) \\
%& = &  
%\alpha(\mathcal{Z}(\tau,\tau,\tau,t)).
%\end{array}
%\]

\end{document}